\documentclass[preprint,12pt,Numbered]{elsarticle} 

\usepackage{graphicx}%
\usepackage{multirow}%
\usepackage{amsmath,amssymb,amsfonts}%
\usepackage{amsthm}%
\usepackage{mathrsfs}%
\usepackage[title]{appendix}%
\usepackage{xcolor}%
\usepackage{textcomp}%
\usepackage{manyfoot}%
\usepackage{booktabs}%
\usepackage{algorithm}%
\usepackage{algorithmicx}%
\usepackage{algpseudocode}%
\usepackage{listings}%
\usepackage{caption}
\usepackage{subcaption}
\usepackage{hyperref}
\usepackage{float}%
\usepackage{pifont}
\newcommand{\cmark}{\ding{51}}%
\newcommand{\xmark}{\ding{55}}%
\usepackage{float}
\restylefloat{table}
\usepackage{caption}

\theoremstyle{thmstyleone}%
\theoremstyle{thmstyletwo}%

\theoremstyle{thmstylethree}%

\journal{}

\begin{document}

\begin{frontmatter}



\title{Generative Manufacturing: A requirements and resource-driven approach to part making}

\author[inst1]{Hongrui Chen\fnref{label2}}
\fntext[label2]{These authors contributed equally to this work.}
\affiliation[inst1]{organization={Carnegie Mellon University},
            addressline={5000 Forbes Ave},
            city={Pittsburgh}, 
            postcode={15213}, 
            state={Pennsylvania},
            country={USA}}

\author[inst1]{Aditya Joglekar\fnref{label2}}
\author[inst1]{Zack Rubinstein}
\author[inst1]{Bradley Schmerl}
\author[inst1]{Gary Fedder}
\author[inst2]{Jan de Nijs}
\author[inst1]{David Garlan}
\author[inst1]{Stephen Smith}
\author[inst1]{Levent Burak Kara\corref{cor1}}
\ead{lkara@cmu.edu}
\cortext[cor1]{Corresponding author}


\affiliation[inst2]{organization={Lockheed Martin Corporation},
            addressline={1 Lockheed Blvd},
            city={Fort Worth},
            postcode={76101},
            state={Texas},
            country={USA}}

\begin{abstract}
Advances in CAD and CAM have enabled engineers and design teams to digitally design parts with unprecedented ease. Software solutions now come with a range of modules for optimizing designs for performance requirements, generating instructions for manufacturing, and digitally tracking the entire process from design to procurement in the form of product life-cycle management tools. However, existing solutions force design teams and corporations to take a primarily serial approach where manufacturing and procurement decisions are largely contingent on design, rather than being an integral part of the design process. In this work, we propose a new approach to part making where design, manufacturing, and supply chain requirements and resources can be jointly considered and optimized. We present the Generative Manufacturing compiler that accepts as input the following: 1) An engineering part requirements specification that includes quantities such as loads, domain envelope, mass, and compliance, 2) A business part requirements specification that includes production volume, cost, and lead time, 3) Contextual knowledge about the current manufacturing state such as availability of relevant manufacturing equipment, materials, and workforce, both locally and through the supply chain. Based on these factors, the compiler generates and evaluates manufacturing process alternatives and the optimal derivative designs that are implied by each process, and enables a user guided iterative exploration of the design space. As part of our initial implementation of this compiler, we demonstrate the effectiveness of our approach on examples of a cantilever beam problem and a rocket engine mount problem and showcase its utility in creating and selecting optimal solutions according to the requirements and resources.\\

\end{abstract}




\end{frontmatter}


\section{Introduction}\label{sec1}
Numerous approaches have been created to facilitate and improve the complex process of part-making. CAD/CAE/CAM software helps in the design, engineering analysis, and manufacturing simulation of a part, while approaches to PLM (Product Lifecycle Management) and PDM (Product Data Management) allows engineering teams and corporations to digitally capture the design and utility of parts and systems and track changes through version control \cite{siemens,autodesk,solidworks,catia, ansys,altair}. However, the original conceptualization of parts is largely performed by humans, focusing most prominently on the engineering requirements. While manufacturing processes and materials may be considered as guidelines and heuristics within DfM (Design for Manufacturing) or, more generally, DfX (Design for X) modules, there is still an unmet need for deploying systems that can take into account the business requirements and prevailing supply chain conditions and accordingly optimize a part. Design, manufacturing, and procurement must be made as seamless as possible to truly optimize a part to the given requirements and resources.

Researchers have begun studying these areas. Recently launched `Generative Design' tools from Autodesk \cite{f360}, nTopology \cite{ntopology}, Altair \cite{altair}, Dassault \cite{dgd}, and several others significantly improve the design optimization process by incorporating various constraints within the optimization that was previously not possible. However, the barrier to the supply chain remains. The design that is output may be feasible but expensive, where this expensive nature of the design can be attributed to the state of the supply chain network (for example, expensive machining equipment and materials). The optimizer can also produce a design that has a large lead time because factors in the supply chain, such as the unavailability of manufacturing equipment required or the need to reorder materials, are not considered by the optimizer. Moreover, trade-offs may exist when comparing different suppliers. In short, there is a need for a system that produces optimal parts, where the topology and other design parameters of each part have been informed by not only the engineering requirements but also the business requirements, such as cost and lead time, which depend on the available suppliers and their capabilities.

\begin{figure*}
\centering

\includegraphics[width=\textwidth]{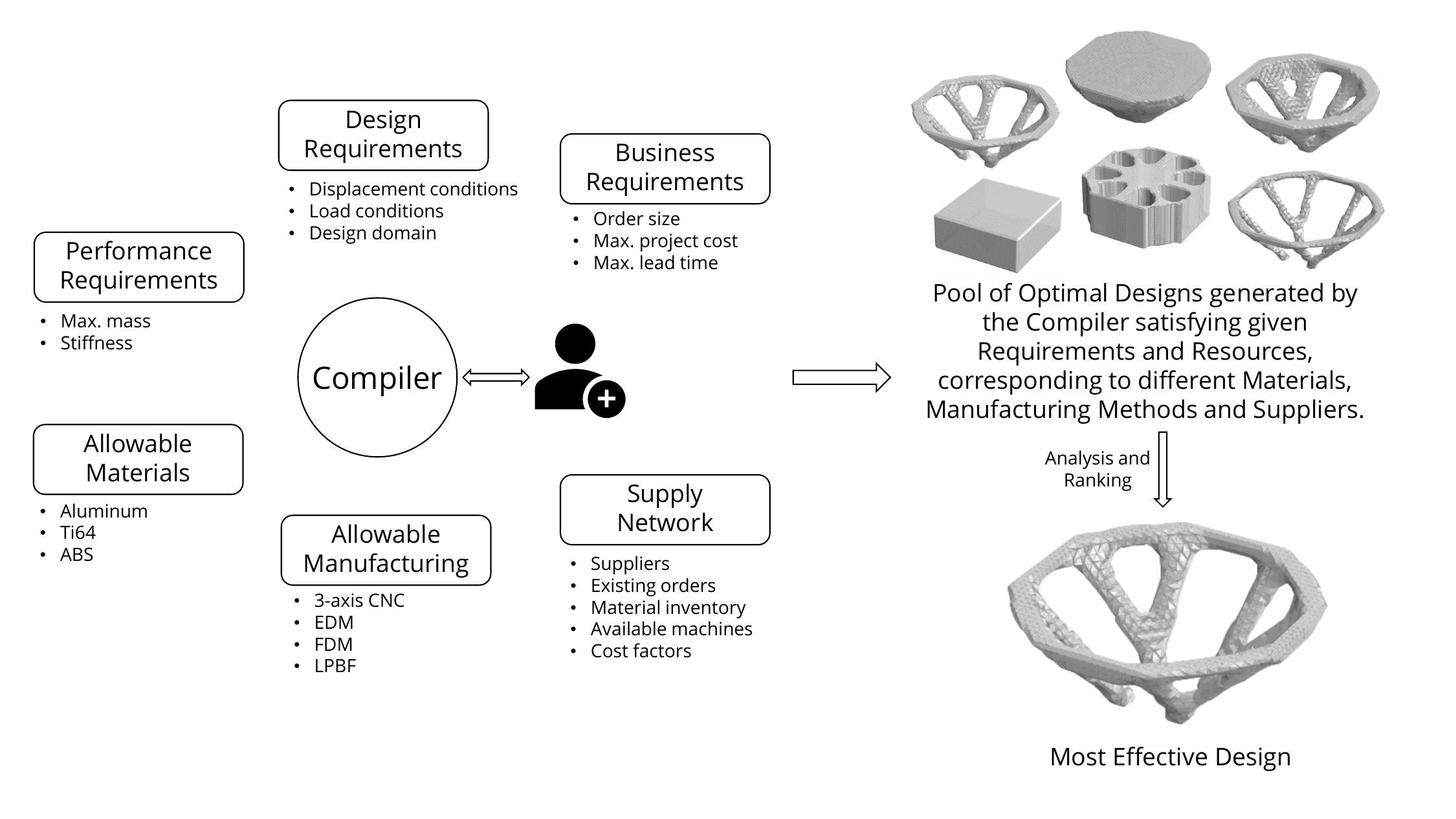}

\caption{Our proposed concept of  Generative Manufacturing: requirements and resource driven design.}

\label{fig:gmconcept}
\end{figure*}

In this work, building towards the goal of removing barriers and combining the design, engineering, manufacturing, and supply chain teams and tasks, we propose a new approach to mechanical part making, which we call Generative Manufacturing (GM) (Figure \ref{fig:gmconcept}). GM enables requirements and resource-driven part-making by informing the part design with real-time supply chain information. In the part-making process for a particular problem, in the first design creation stage itself, our system can provide answers to several questions: 
\begin{enumerate}
    \item Which manufacturing method (e.g., additive or subtractive) will result in the shortest lead time for the product?
    \item What constraints are active (impacting solution) vs. inactive (not influential)?
    \item Why a particular manufacturing method (e.g., 3-axis CNC) is infeasible given the constraints?
    \item What are the trade-offs between choosing different materials (e.g., Al6061 and Ti6Al4V) for this product?
    \item How will the best solution change if the mass and cost constraints become stricter?
    \item How will the best solution change if the state of the suppliers changes? 
\end{enumerate}

Our approach works by first probing the current manufacturing state of each potential supplier for a given set of part designs to create models that estimate the current relationships between parts, cost, and lead time for each supplier. This is accomplished via a distributed framework that situates a finite capacity scheduler at each supplier site and utilizes current knowledge of previously accepted supplier orders, machine availability, and material inventories to project supplier-specific cost and lead-time for particular part design requests. At the heart of the approach is a neural network-based differentiable design generator that can incorporate these supply network models as well as other requirements and resources that it receives as input for creating optimized part designs. As part of our initial implementation of these ideas, we demonstrate our approach in the design of a rocket engine mount and a cantilever beam. We showcase different requirements and resources with different supply chain scenarios and how they inform the part design. We show that our approach enables a user-guided iterative exploration of the design space where requirements can naturally evolve in response to design ideas suggested by our system. By providing a portfolio of competitive but often times surprising solutions to a given problem, our method also helps end-users ‘discover’ requirements that were either ill-posed or under-constrained, bringing design optimization closer to a dialogue between human and the machine rather than treating generative part-making as the black-box solution to an optimization problem.
Our main contributions include:
\begin{itemize}
\item{Presenting the concept of generative manufacturing: requirements and resource driven part making.}
\item A design generator that performs topology optimization with manufacturing, time, and cost constraints.
 \item An incremental, finite-capacity scheduler that imports constraints characterizing the current manufacturing state of a given supplier and uses this knowledge to produce manufacturing cost and lead-time options for the design generator to bias future part design decisions.
 \item An interactive tool for exploring critical decision variables and thresholds to understand the design space of generated manufacturing options.
 
\end{itemize}

The implementation of this work is available at \url{https://github.com/AdityaJoglekar/Generative_Manufacturing}.

\section{Background}\label{sec2}
\subsection{Generative Design and Manufacturing}

The primary motivation behind embracing generative design systems involves leveraging computational power to assist human designers and potentially automate aspects of the design process. Alongside achieving efficiency, cost savings, optimization, accuracy, and consistency, an essential goal is to expand exploration within the design realm and facilitate design creation \cite{singh2012towards}. On the manufacturing side, the industrial Internet of Things devices consist of manufacturing data that feeds into the operation decision \cite{daniels2023building}.  It is important to design a framework to integrate design and operation decisions\cite{van2023digital}. One of the main claimed advantages of generative design is the incorporation of constraints and offering design candidates \cite{f360,ntopology,altair,dgd}. However, at its core, generative design relies on topology optimization to satisfy the structural performance. Topology optimization approaches such as SIMP (Solid Isotropic Material with Penalisation) (\cite{bendsoe1989optimal,zhou1991coc}) and the level-set method (\cite{allaire2002level}) help solve the highly complex and non-convex problem of optimum material layout for different engineering objectives. Several extensions, both in research \cite{morris2020subtractive,Qian2017,Mirzendehdel2016,mirzendehdel2020topology} and commercial software \cite{f360,ntopology,altair,dgd}, to include manufacturing constraints have also seen success. To incorporate business requirements like lead time and cost into the optimization process, it is essential to use computational models that estimate these quantities.

While there exist supply chain-dependent lead time and cost estimation models \cite{apriori}, we were not able to find frameworks that integrate them into design optimization. Our proposed framework offers a first step in this direction. Extensive work has been done on creating theoretical and empirical models to predict the manufacturing time and cost, given the features of the part. Noting that our final goal is to integrate the model with topology optimization, we require a computationally light model (as this model will need to be called during the optimization iterations) and a differentiable model (gradients to inform the design need to exist). These criteria rule out CAM simulation software as a possible model. While empirical models, such as using supervised machine learning for prediction, can be very effective, they require a large amount of data, in the absence of which they can face generalization issues. Different parametric models have been developed that are computationally inexpensive and can be considered sufficiently accurate, especially for the design optimization phase. This indicates their suitability over other models for achieving our goal. In the next subsection, we review the related parametric models for additive and subtractive manufacturing methods.

\subsubsection{Additive manufacturing oriented topology optimization}
The need for integrating additive manufacturing considerations in topology optimization is highlighted in \cite{seepersad2014challenges}. A substantial body of research has delved into the detection and mitigation of overhang edges to minimize the need for support structures \cite{Thompson2016,Leary2014,Brackett2011,Gaynor2014,Mhapsekar2018,Langelaar2016,van2018continuous,van2018continuous,Mirzendehdel2016,Liu2017Support,Zhang2019,Qian2017,barclift2017cad}. Qian et al. \cite{Qian2017} employed linear interpolation between nodes of a finite element mesh to achieve a more accurate density gradient. The incorporation of density gradient has facilitated the inclusion of self-supporting structures, boundary slope control, and print angle optimization for simultaneous optimization with topology \cite{Mezzadri2018,Wang2019,Wang2020}. Our overhang detection method builds upon the work of Wang and Qian and the work by Chen et al. \cite{chen2023concurrent}, integrating the print angle through the vector dot product of the print angle with the filtered density gradient. This method distinguishes itself with (1) an accurate and differentiable density gradient derived directly from the neural network, enabling topology optimization without the need for filtering, and (2) support structure modeling from the overhang.

\subsubsection{Subtractive manufacturing oriented topology optimization}
Subtractive machining is a widely used method in manufacturing. Subtractive machining refers to manipulating the cutting tool to remove material until the desired geometry is reached. Two instances of subtractive machining include milling and 2D cutting. The tool head can be manipulated in 3, 4, or 5 degrees of freedom in milling operation. Whereas 2D cutting methods (laser, water jet, electrical discharge machining) perform through cut and the final design can be considered a 2D extrusion of a shape profile. Langelaar \cite{langelaar2019topology} proposed a machining filter to optimize topology for the multi-axis machining process. The machining filter can be applied to 2.5D and 4-axis machining. We adopt the machining filter to 3-axis machining where the user can specify a combination of 6 possible machining orientations along the principal axis. Other prior works related to subtractive machining include projection-based approach \cite{vatanabe2016topology,guest2012casting}, feature-based approach with level set \cite{liu20153d}. Further adaption of projection-based approach has been applied to casting as well \cite{gersborg2011explicit}. On the other hand, 2D cutting can be directly formulated as a 3D optimization of a 2D profile extrusion, and a neural network-based direct topology optimization can be developed to do so \cite{Chandrasekhar2021}.

\subsection{Supply Chain Scheduling}

Manufacturing scheduling problems have been extensively studied for over 60 years \cite{Morton1993, Pinedo2016}. Traditionally, they have been approached within the Operations Research community through the use of mathematical modeling techniques (c.f., \cite{Nemhouser1994}), but more recent advances in constraint reasoning and heuristic search from the AI community (e.g., \cite{Smith1987, Laborie2009}) have also established the power of constraint-based search and optimization techniques as practical solving techniques for this class of problems. Consideration of the broader scoped problem of supply chain scheduling has a much more recent history. As summarized in \cite{Chen-Hall2022}, this merger of two disciplines - scheduling and supply chain management - focuses principally on solution of larger coupled optimization problems (e.g., integrated production and distribution scheduling, joint scheduling, and supplier pricing) and on coordinated decision-making by multiple decision makers (in both centralized and decentralized settings). However, business enterprises have been slow to exploit this more recent research. Enterprise Resource Planning (ERP) systems, which are fundamentally driven by estimates of predicted performance and often have little connection to the enterprise's actual current manufacturing state, still dominate the operational landscape and present an important challenge to the goal of generative manufacturing. Our approach in this paper is to exploit available information on currently booked orders, material inventories and replenishment constraints, and machine maintenance requirements (much of which is already available in existing ERP systems) to enable accurate projection of current cost and lead time for taking on a new manufacturing request.

\subsection{Explainability for manufacturing}
With the increase in options that a designer has in a DfX context, it becomes difficult for the designer to understand and explore the nuances of the design space. In particular, the designer needs to understand the following aspects of the design space: (1) the tradeoffs that are happening between the different qualities of a design (cost, lead time, rigidity, strength, etc.) and how they are related, and (2) which of these qualities have the most impact on the design. 

Design space exploration has been broadly studied in many areas of software, including product lines~\cite{murashkin2013visualization}, model-based performance prediction~\cite{DBLP:journals/tse/BalsamoMIS04} and formal verification~\cite{10.1007/978-3-642-21292-5_3}. However, {\it explanation of design spaces} remains a challenge. Recent work~\cite{DBLP:conf/ecsa/CamaraSGS21,DBLP:journals/jss/CamaraWGS23,DBLP:journals/jss/WohlrabCGS23,ieeesoft23} has explored the use of dimensionality reduction techniques, traditionally used in areas such as biology and machine learning~\cite{pcanature}, to identify and facilitate the understanding to a human designer of the main design decisions and tradeoffs in a design space.

In~\cite{DBLP:journals/jss/CamaraWGS23,DBLP:journals/jss/WohlrabCGS23}, 
Decision Tree Learning (DTL)~\cite{Breiman2017} is used to explain how concrete choices associated with specific design decisions influence the qualities across the design space (among other techniques).
These techniques are generalized in~\cite{ieeesoft23}, which describes a design space explanation process and lessons learned from experience in those domains, as well as the generative manufacturing case.

\section{Proposed Solution: Generative Manufacturing}\label{sec3}
\begin{figure*}
\centering

\includegraphics[width=\textwidth]{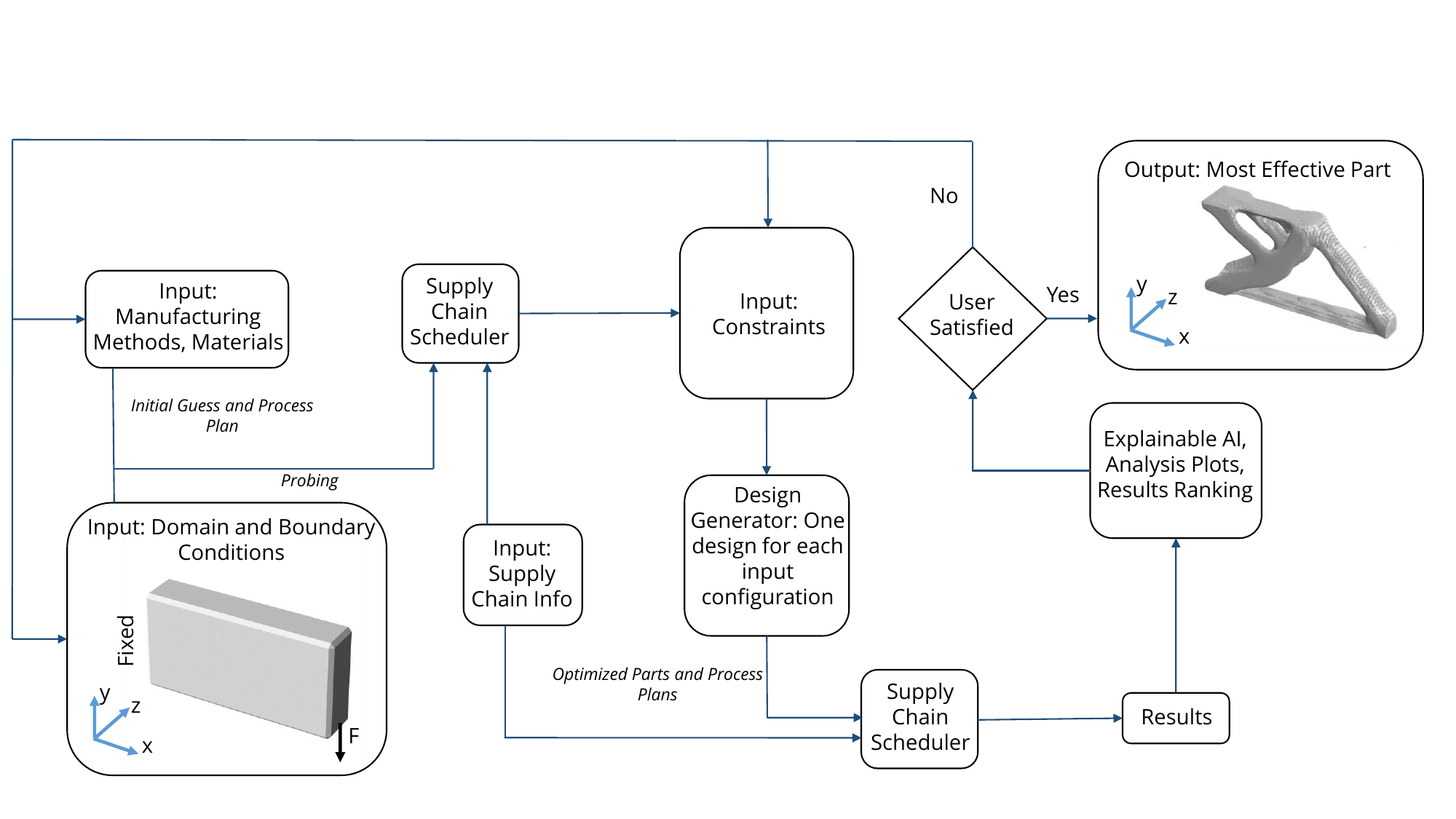}

\caption{Our proposed framework. 1) The user inputs the problem domain and boundary conditions and a set of manufacturing method and material combinations. 2) An initial probe of the supply chain using system generated representative part guesses helps swift removal of infeasible combinations. This process also helps establish the relationship between different requirements (mass, compliance, lead time, cost) for each of the current suppliers, gives an approximate range of values for each of the requirements that help the user determine the constraints for mass, lead time and cost for performing topology optimization in the design generator and also helps determine the best supplier to achieve Pareto optimal solutions. 3) One optimized design corresponding to each input configuration (defined by boundary conditions, manufacturing method, material and supplier) is output by the design generator and then passed through the supply chain scheduler to get the lead time and cost. 4) All the designs are then evaluated and visualized using the Explainable AI and Results Interface, where trade-offs are explored, which helps in user feedback to the system and selection of the most effective part.}

\label{fig:flowchart}
\end{figure*}
For generating the most effective part possible given the requirements and resources, we propose a system that integrates a novel Design Generator, Supply Chain Scheduler, and an Explainable AI and Results Interface as shown in Figure \ref{fig:flowchart}. The engineering domain and boundary conditions, mass, structural rigidity, lead time, cost, materials, manufacturing methods, and suppliers are the engineering and business requirements and resources we consider in our current system. Note that our system is flexible enough to include other requirements, such as maximum stress or thermal considerations, but we leave its implementation for future work. Details of the system modules are presented in the following sections.


\subsection{Supply Network Scheduler}

To evaluate the cost and lead time implications of a candidate design in light of the current manufacturing state of the supply network, the system incorporates a supply-side scheduler. The scheduler takes as input from the design generator a manufacturing request consisting of a part type, the quantity required, the date by which the manufactured parts are needed, and a set of candidate designs with associated process plans. It also receives information from each supplier capable of producing the candidate part designs relating to the supplier's current operating state, including the existing set of accepted orders, the types and numbers of manufacturing machines and processes available along with their operating characteristics and costs, current material inventories and costs, and other availability constraints. To address privacy concerns with respect to supplier business information, candidate designs are evaluated in a decentralized manner, where an instance of the scheduler is situated with each supplier, and the design generator independently queries each supplier to obtain cost and lead time estimates, which are formulated as bids, for a given input request and set of candidate designs. Subsequent analysis of the set of bids returned is then used to adjust constraints for the next iteration of design candidates.

Upon receipt of a new request, the scheduler produces bids for its associated supplier by generating a finite capacity production schedule that includes both the existing set of accepted orders, which are either in-process in the factory or planned, and the new request. The scheduler, which is designed to accept and schedule requests incrementally over time, first allocates machine capacity and materials required to execute the process plans associated with all existing orders over time while respecting any other known constraints on resource availability, which can include planned downtime for machine maintenance and material resupply times. Each process plan specifies a sequence of manufacturing tasks, such as `printing $\to$ sintering $\to$ ...', with each task designating its required capabilities, such as 3-axis machining, its nominal duration, and its nominal cost. Once this ``current'' schedule has been created, it is extended to include the process plan associated with the new request, utilizing whatever available machine and material capacity remains over the scheduling horizon. This hypothetical schedule is then used to provide the lead time and cost estimates for this supplier bid. 
For those input requests that provide multiple part design options, a different hypothetical production schedule is generated for each corresponding process plan, and separate bids are returned for each option.

\subsubsection{Basic Generation of Supplier Options}

In basic bid generation mode, the scheduler attempts to integrate the tasks associated with manufacturing a candidate design into the production schedule so as to minimize lead time, subject to the constraint that existing accepted orders have priority and are not delayed to accommodate this ``due date quote''. The associated process plan is ``instantiated'' by the scheduler to create a network of tasks, splitting the total number of parts ordered into a set of manufacturing lots that can be produced in parallel if sufficient manufacturing resources, which include the machines and materials, are available. Each instantiated task is then interrogated to determine the set of supplier machines that could be used to carry out this manufacturing step. Given the determined machine alternatives and the design's material requirement, a search is performed to determine the choice of machine assignments to tasks that yield the best result. In this case, it is the set of assignments that produces the minimal lead time. If there is insufficient material on-hand to produce the full quantity of parts requested, then the scheduler adds a resupply time constraint to delay production of the candidate design option until the material required to produce it is available. 

To ensure the feasibility of any given assignment of machines to tasks, the scheduler relies on an underlying graph of time points and edge-weighted distances called a Simple Temporal Network (STN) \cite{Dechter71}. The start and end points of all tasks included in the schedule are encoded in the STN, as are the sequencing constraints dictated by order process plans, the constraints introduced by the scheduler to serialize tasks that have been assigned the same machine, and any other availability constraints that must be taken into account such as material resupply time. To generate an assignment for all tasks in the instantiated task network, the search proceeds to consider each task in the instantiated task network in topological order. At each step, the search moves forward through the sequence of tasks currently assigned to each resource capable of performing the next unscheduled task, which is referred to as each resource's current timeline, looking for temporal gaps large enough to accommodate this task. As each temporal gap is tested, constraints are propagated in the underlying STN to confirm the continued feasibility of this partial schedule or to signal conflict and the need to move on to the next temporal gap. When a feasible assignment is found for all tasks in the instantiated network, its objective score, which in this case is the instantiated task network's overall scheduled end time, is recorded, and the search moves on to consider alternative resource assignments. When the search is completed, the feasible assignment with the earliest overall end time (i.e., the smallest lead time) is selected as the basis for generating the bid option.

The lead time and cost estimates reflected in this generated schedule are biased by supplier-specific refinements to the nominal duration and cost values specified in the candidate design's input process plan. The scheduler operates with a model of the supplier's actual resources (machines and materials) that include coefficients for tuning nominal task durations and costs to the characteristics of the supplier's specific assets as well as the supplier's specific pricing procedures. A supplier may have multiple instances of a particular 3D printer, for example, but they may range in age and consequently have different operating speeds. Similarly, the wear and tear on a milling machine as well as the milling time required to achieve a particular part geometry will vary as a function of the density of the material, and the magnitude of cost coefficients capture supplier-specific operating costs and price margins.

\subsubsection{Utilizing Multiple Suppliers through Combinatorial Auction}

\begin{figure}
\centering

\includegraphics[width=3.2in]{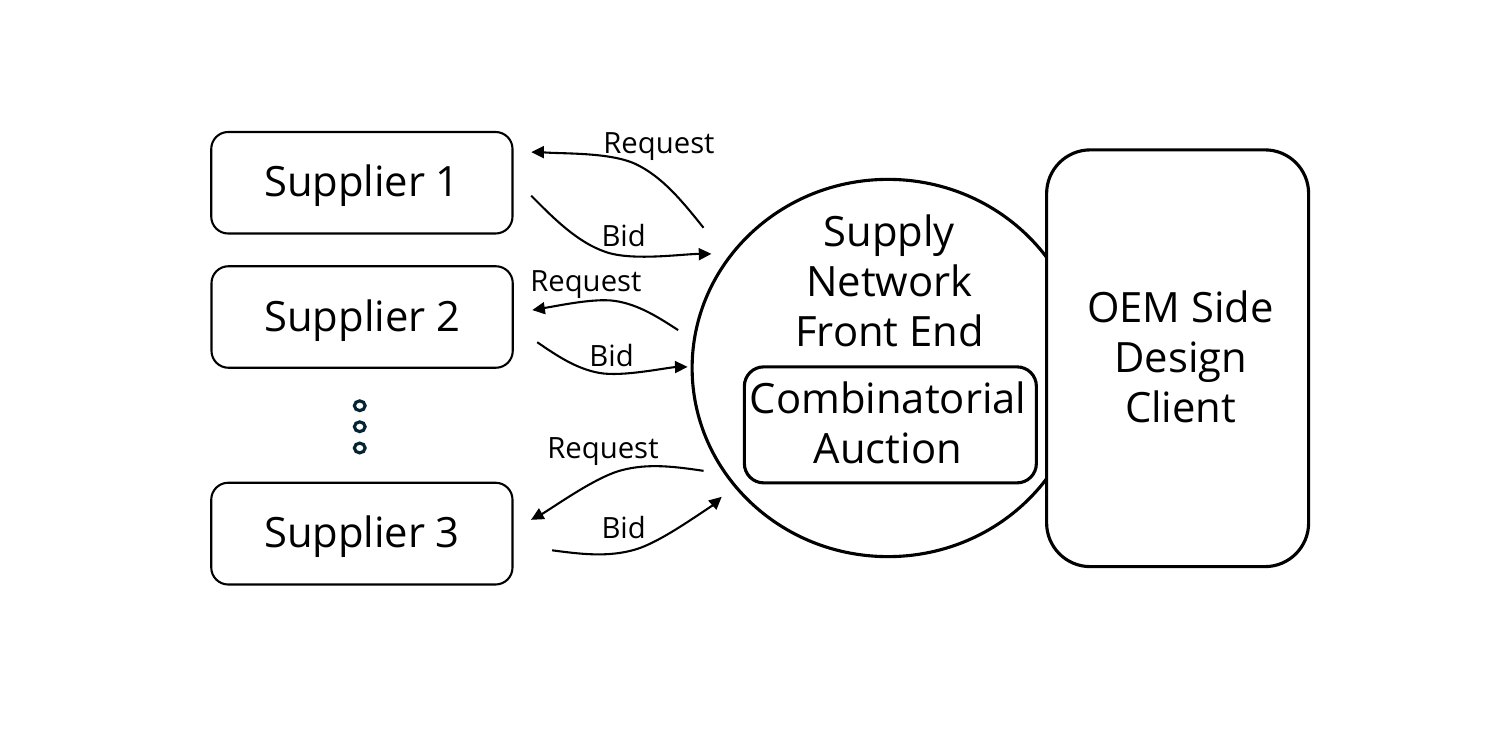}

\caption{Decentralized Coordination with Supply Network}
\label{fig:comb-auction}
\end{figure}

When the scheduler is operating in basic bid generation mode, it is assumed that each supplier will generate independent bids for manufacturing a given candidate design. However, for requests with large part quantities this may not be feasible or practical. To accommodate such situations, the scheduler can also be configured to treat the request's `needed by' date as a hard constraint and instead generate partial bids that indicate the number of parts that can be produced while meeting this constraint. In this mode, partial bids generated by different suppliers are assembled into complete multi-supplier bids through application of a combinatorial auction and the resulting complete bids are passed on to the manufacturer-side design client as before for analysis and feedback to the design generator. The combinatorial auction employs a search process that can be configured to emphasize different criteria for determining how to best combine the partial bids of different suppliers, such as minimizing the number of suppliers and producing the lowest cost bid. Figure \ref{fig:comb-auction} illustrates the overall decentralized framework for querying the supply network's current capability to handle requests to manufacture quantities of parts according to various candidate designs.

\subsection{Design Generator}
\begin{figure*}
\centering

\includegraphics[width=\textwidth]{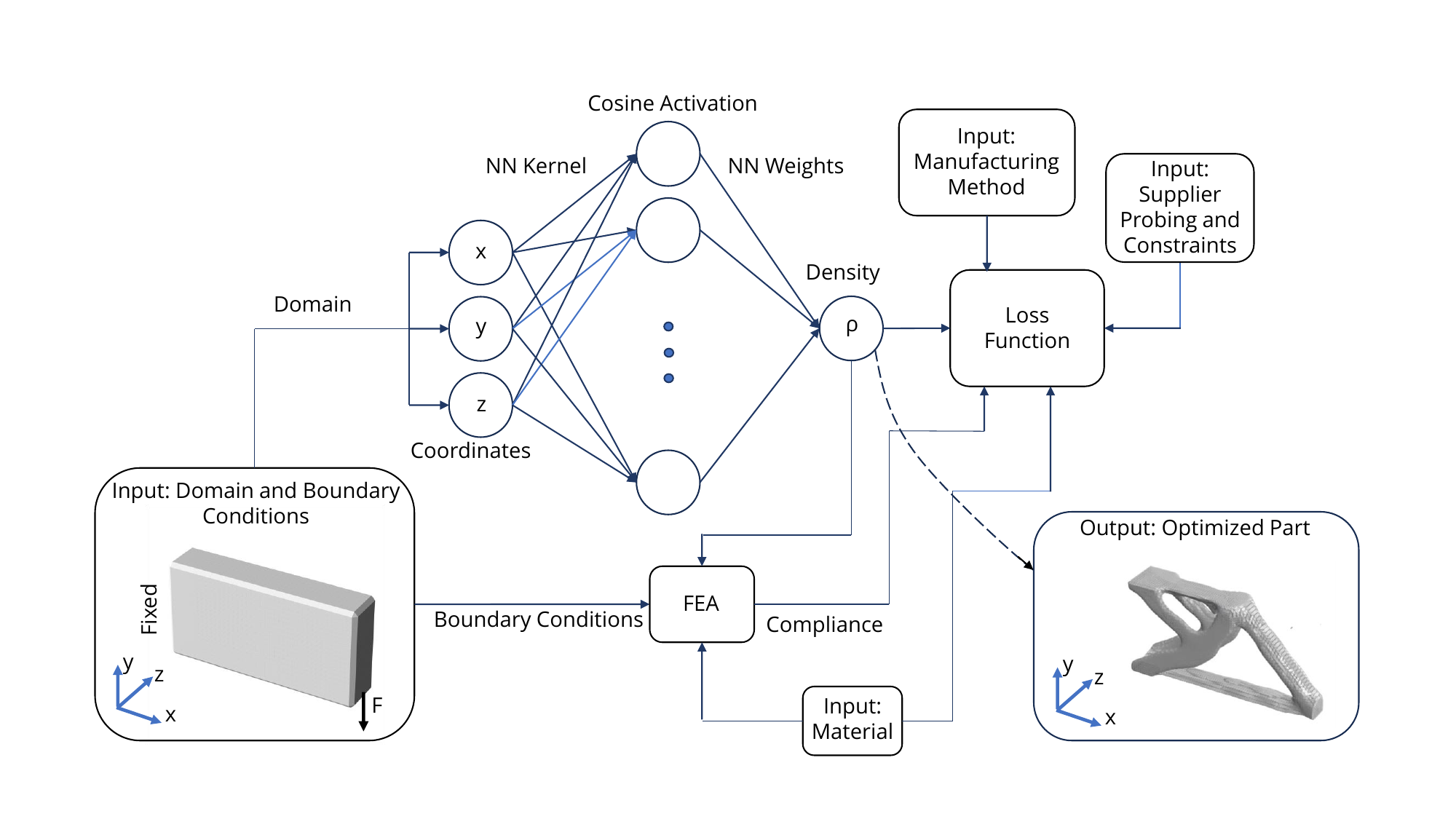}

\caption{The design generator. 1) The problem domain defines the coordinates input into the neural network (NN) that outputs the density at each of these coordinates. 2) The density values define the topology of the part on which the quantities in the loss function depend. 3) The material, manufacturing method, supplier probing and constraints help define the loss function. 4) The NN learns the weights (using backpropagation and gradient descent) to output the optimal topology that minimizes the loss function.}

\label{fig:designer_flowchart}
\end{figure*}

The design generator takes in as inputs the engineering domain and boundary conditions, manufacturing method, material, supply chain situation and constraints on mass, lead time and cost, and performs topology optimization with an objective of minimization of compliance and outputs the optimized part, as shown in Figure \ref{fig:designer_flowchart}.

We utilize a neural network for representing the part geometry (the continuous density distribution field within the domain) which gives us the ability to easily optimize functions of the part boundary and its gradients. This is particularly useful for the cost and time objectives as they are dependent on these quantities (details in section \ref{loss_function_formulation}).

Manufacturing methods and materials are inherently not continuous variables to optimize over. There exist techniques for continuous approximations of materials, such as considering the Young's Modulus as the continuous variable representing the material. However, these approximations are often inaccurate and, in our case, would result in an unwarranted increase in the complexity of the already complex and highly non-convex optimization problem. Moreover, our system is designed such that starting from a diverse but finite set of material choices, the user can focus on a particular set of materials as the iterations progress. Hence, we keep the materials as a discrete variable in the overall optimization process. Manufacturing methods are not related to each other, exist as separate entities in space, and are finite in amount. Also, similar to materials, the user can focus on certain manufacturing methods as the iterations progress. Hence, manufacturing methods are also considered as discrete variables in our optimization setting. 

The lead time and cost depend on the prevailing supply chain situation, and topology optimization with these objectives is challenging. Modeling of a differentiable function that maps the topology to these quantities is required. We use `supplier probing' for this. Canonical forms, or guesses, that depend on the engineering domain and boundary conditions, the manufacturing method, and the material are created. These guesses span many volume fractions within the domain. Creation of these guesses requires extremely low computation compared to the topology optimization of a part. We use differentiable and efficient parametric models for finding the nominal time and nominal cost for each of these guesses. Then, these guesses are sent to the supply chain scheduler to find the corresponding lead time and cost. Using regression, we find the current relationship that exists between the volume fractions, nominal values of time and cost, and supplier values of time and cost for each combination of material and manufacturing method for each of the suppliers. Now, any user given constraints on lead time and cost can be used to find corresponding nominal time and nominal cost, for which differentiable mappings to the topology exist. Thus, we can achieve topology optimization that satisfies the lead time and cost constraints. We provide the details of this process for different manufacturing methods in section \ref{loss_function_formulation}.

The easiest method for a user to prescribe the requirements such as mass, cost, and lead time is to specify values constraining these quantities. Hence, we use a concept similar to the epsilon-constraint method, which is one of the primary methods of solving multi-objective optimization problems. We formulate the optimization problem with four objectives of compliance, mass, cost, and lead time, and minimize compliance with user-given constraints on mass, cost, and lead time to obtain a Pareto optimal solution. The optimization can also be performed with a different combination of the above four terms corresponding to minimization and constraints. The supplier probing we perform guides the user in specifying the values of these constraints by getting an estimate of the possible lowest and highest values. After the optimization for the current set of requirements and resources is complete, the analysis plots and decision trees help the user understand the trade-offs and helps the user effectively change the constraint values if required and perform another iteration of the optimization with these new requirements. Note that whenever the requirements and resources change, our system performs the topology optimization for creating the set of optimal solutions corresponding to these new inputs. The density value at each coordinate of the parts produced by our system is dependent on, informed by, and optimal in terms of all these requirements and resources.

We describe the density neural network first, which forms the basis of our design generator, and then explain in detail the loss function formulation for different manufacturing methods.

\subsubsection{Density Neural Network}\label{density_neural_network}
The density neural network $\textit{Den}(\mathbf{X}_{den})$ can be represented as follows:
\begin{equation}
    \textit{Den}(\mathbf{X}_{den}) = \sigma((\cos(\mathbf{X}_{den}\mathbf{K}_{den} +  \mathbf{b_{1}})+\mathbf{b_{2}})\mathbf{W}_{den} + \mathbf{o_{1}})
    \label{eq:dnn}
\end{equation}
The input is a batch of domain coordinates $\mathbf{X}_{den(\text{batchsize} \times 3)}$. We use the domain center as the origin for the coordinates, and the coordinates are normalized with the longest dimension coordinates ranging from -0.5 to 0.5. We use the concepts proposed in \citet{tancik2020fourier} and \citet{sitzmann2020implicit} and a neural network architecture similar to the one used in \citet{Chandrasekhar2021Fourier} and \citet{chen2023concurrent}. The first layer weights (kernel $\mathbf{K}_{den(3 \times \text{kernelsize})}$) are fixed, which creates Fourier features after passing through the cosine activation. The kernel is created using a grid of a number of dimensions the same as the number of domain dimensions, and then reshaping the grid coordinates to the matrix $\mathbf{K}_{den(3 \times \text{kernelsize})}$. The grid size in each dimension dictates how well it can represent topological features, and the grid's range of values controls the frequency of the output topology, with higher ranges of values giving a topology with more intricate features. Trainable biases ($\mathbf{b_{1}}$ and $\mathbf{b_{2}}$) are added to improve the expressive power of the neural network. The next layer weights ($\mathbf{W}_{den(\text{kernelsize} \times 1)}$) are trainable. This output is passed through a sigmoid activation ($\sigma$), that ensures final output values are between 0 and 1, which represent the density, for each of the coordinates in the input batch. We find empirically that the best initialization of the neural network is such that a uniform density topology with volume fraction corresponding to expected active constraint is output. We achieve this by setting $\mathbf{W}_{den(\text{kernelsize} \times 1)}$ close to zero and adding an appropriate offset ($\mathbf{o_{1}}$) before applying the sigmoid activation (details of offset calculation in section \ref{loss_function_formulation}). The density distribution output by the neural network is used to calculate the different terms in the loss function of the neural network. We use Adam (\citet{kingma2014adam}) as the optimizer, with a learning rate of $2.0\times10^{-3}$ for all the experiments.

\subsubsection{Loss Function Formulation}\label{loss_function_formulation}

Each manufacturing method has particular characteristics that define the constraints on the part topology that can be created, as well as the cost and time for manufacturing the topology. We utilize and build upon existing works to achieve requirements and resource-driven topology optimization for additive and subtractive manufacturing. In additive manufacturing, we consider LPBF for metals and FDM for plastics, and in subtractive manufacturing, we consider 3-axis milling and 2-axis cutting with EDM. We present the detailed loss function formulation for each of these manufacturing methods in this section. Our framework is extensible to different manufacturing methods, but we limit the scope of this paper to only the ones mentioned above.

\paragraph{Additive Manufacturing}\label{Additive}~\\
The manufacturing process considered is as follows:\\
$\text{Machine setup} \rightarrow \text{Printing} \rightarrow \text{Support Removal} \rightarrow \text{Inspection}$\\
Objective Function:\\
\begin{equation}
\begin{split}
\text{Loss} = &\frac{c}{c_0} + \alpha({\text{max}(0,\frac{\text{mass}}{\text{masscon}} - 1.0)}^2) +\\ &\alpha({\text{max}(0,\frac{\text{cost}}{\text{costcon}} - 1.0)}^2) + \alpha({\text{max}(0,\frac{\text{time}}{\text{timecon}} - 1.0)}^2)     
\end{split}
\label{eq:addobj}
\end{equation}\\
where, the notation definitions are as given in Table \ref{tab:obj}.
\begin{table}[H]
\begin{tabular}{l|l}
     $c$ &compliance of current topology  \\
     $c_0$ &compliance normalization constant\\
     $\text{mass}$ &mass of current topology\\
     $\text{masscon}$ &mass inequality constraint (mass $\leq$ masscon)\\
     $\text{cost}$ &cost of current topology\\
     $\text{costcon}$ &cost inequality constraint (cost $\leq$ costcon)\\
     $\text{time}$ &time of current topology\\
     $\text{timecon}$ &time inequality constraint (time $\leq$ timecon)\\
     $\alpha$ &penalty coefficient
\end{tabular}
\caption{Notation definitions}
\label{tab:obj}
\end{table}

Note that each of the variables in the above loss function, i.e. the compliance, mass, cost, and time, must be a differentiable function of the density values at each of the coordinates for the topology to be optimized with respect to them.

The compliance ($c$) can be formulated as such as shown in \cite{bendsoe1989optimal,zhou1991coc} using the SIMP method and used in a self-supervised neural network topology optimization approach as shown in \cite{Chandrasekhar2021Fourier, chen2023concurrent}.

The volume fraction (vf) of a part discretized into $n$ elements for the SIMP method, with $\rho_i$ being the density value at each of these elements, is defined as follows:\\
\begin{equation}
        \text{vf} = \frac{\sum_i^n \rho_i}{n}
    \label{eq:vf}
\end{equation}
The mass can be easily formulated as a function of $\rho_i$ as follows:
\begin{equation}
    \text{mass} = \sum_i^n \rho_i  v  d
    \label{eq:mass}
\end{equation}
where $v$ is the unit voxel volume and $d$ is the material density.

We use the following parametric equation for defining the nominal time ($t_{nAM}$) in terms of the density values:
\begin{equation}
     t_{nAM}= t_{pAM} + t_{sAM} + t_{rAM} + t_{iAM}
\label{eq:addtime}
\end{equation}
where $t_{pAM}$ is the print time, $t_{sAM}$ is the setup time, $t_{rAM}$ is the support removal time and $t_{iAM}$ is the inspection time.
The print time is calculated as follows:

Firstly, the support structure volume is calculated, wherein the overhang region ($P$) is found using the differentiable nature of the neural network as shown in \cite{ chen2023concurrent}. $P$ is a tensor of the same shape as xPhys, where $\text{xPhys} = \textit{Den}(\mathbf{X}_{den})$. Then, we use cumulative summation along the print axis to get $Pcs$, followed by a Heavyside function to get $Ph = \frac{1}{(1+ \exp^{-10(Pcs-2)})}$. We then perform an element-wise product of $Ph$ with (1-xPhys) to get $Pv$ and perform a summation of element values of $Pv$ to calculate the exact support structure volume. The support structure mass is then calculated using Equation \ref{eq:mass}, where the $\rho_i$ now corresponds to the support structure. A support structure material density of $k$ times the material density is used (we use the common value of k = 0.3 for all the results). The part mass is calculated using Equation \ref{eq:mass}. Now, print time is:
\begin{equation}
t_{pAM} = \frac{m_{part} + m_{support}}{Q_{AM}}
\end{equation}
where $m_{part}$ is the part mass, $m_{support}$ is the support structure mass and $Q_{AM}$ is the print rate.

The print rate depends on the material being used and can be input by the user. We use standard values for each material in all the examples in section \ref{Results}. Also, we currently use constants to denote the values of setup time, support removal time, and inspection time for all the examples in section \ref{Results} (we use Trumpf-TruPrint 3000 as a reference) (refer to \url{https://github.com/AdityaJoglekar/Generative_Manufacturing} for all the standard values and constants used). We believe Equation \ref{eq:addtime} is a good approximation for our use case and leave using more complex nominal time equations for future work.

For the nominal cost ($c_{nAM}$), we use the following equation:
\begin{equation}
     c_{nAM}= t_{pAM} \times c_{pAM} + c_{mAM} + c_{sAM} + c_{rAM} + c_{iAM}
\label{eq:addcost}
\end{equation}
where $t_{pAM}$ is the printing time (in minutes), $c_{pAM}$ is the printing cost per minute, $c_{mAM}$ is the material cost, $c_{sAM}$ is the setup cost, $c_{rAM}$ is the support removal cost and $c_{iAM}$ is the inspection cost.

The printing cost per minute is a standard value we input. The material cost is the total printing mass in kg times the material cost per kg, where again we use standard material cost values in the examples shown in section \ref{Results}. The remaining terms in the nominal cost equation are considered standard constant values similar to the nominal time equation. Note that the supply chain model modifies all these standard values according to each supplier's capabilities and resources. They impact the topology as shown in the following example: for the same constraint on cost, a higher setup cost constant value will lead to lower print time possible and thus lower volume and mass possible for the topology.

Now, given the actual cost and lead time constraints, the probing procedure will help determine the corresponding volume fractions and constraint values to be used in the loss function.

Probing: We generate 13 representative topologies corresponding to volume fractions (vf) ranging from 1.0 to 0.005 (we found empirically this works for a large number of problems), where each topology consists of all elements of $\rho_i = \text{vf}$, and we assume support volume equals to $0.1\times\text{vf}\times\text{total volume of design domain}$. Mass, nominal time, and nominal cost are calculated and process plans are generated for each of these topologies and passed to the supply chain model to get the actual cost and lead time. We utilize the fact that the volume fractions and the mass, actual cost, and lead time are highly correlated and create linear regression models using these 13 topologies. Each model has a volume fraction as the input and the actual cost or lead time corresponding to a supplier as the output. Hence, given a lead time or actual cost constraint by the user, we can map this constraint to a volume fraction. Note that here indirectly a mapping between the nominal time and lead time is also occurring as each nominal time corresponds to a volume fraction (similarly for nominal cost and actual cost). Hence, given the actual cost and lead time constraints, we can input the corresponding nominal cost and nominal time values as costcon and timecon in Equation \ref{eq:addobj} and use Equations \ref{eq:mass}, \ref{eq:addtime} and \ref{eq:addcost} in Equation \ref{eq:addobj} to get a differentiable objective function that can change the topology with respect to the supplier.

Probing also helps in finding the approximate active constraint. For additive manufacturing, we can do so by finding the constraint that has the lowest corresponding volume fraction ($minvf$). For the mass constraint, the corresponding volume fraction can be easily found by using Equations \ref{eq:vf} and \ref{eq:mass}, and for actual cost and lead time constraints, probing can be used as described above. For compliance minimization, the highest volume fraction is the best, but the active constraint will be violated if the volume fraction of the optimized topology goes any higher than $minvf$. We can use this fact about the active constraint for initialization of the neural network for reasons explained in section \ref{density_neural_network}. The offset $\mathbf{o_1} = \text{log}(\frac{minvf}{1-minvf})$ is used in Equation \ref{eq:dnn}. We also find empirically that setting the compliance normalization constant ($c_0$) in Equation \ref{eq:addobj} to the compliance corresponding to a topology with uniform density and volume fraction equal to $minvf$ gives the best results.

For the penalty coefficient $\alpha$ in Equation \ref{eq:addobj}, we find empirically that using the following schedule gives the best results: Initialize $\alpha = 0$ and increment by 0.5 in each optimization iteration until 100 iterations. Then increment by $(\frac{\text{iteration number}}{100})^3$ until $\alpha = 100$, and keep $\alpha  = 100$ for remaining iterations.

\paragraph{Subtractive Manufacturing}\label{Subtractive}~\\
\textbf{3-axis milling:}\\
The manufacturing process considered is as follows:\\
$\text{Machine setup} \rightarrow \underbrace{\text{Fixture setup} \rightarrow \text{Machining Operation}}_{\text{n times}} \rightarrow \text{Polishing} \rightarrow \text{Inspection}$\\
Objective Function:\\
\begin{equation}
\begin{split}
\text{Loss} = &\frac{c}{c_0} + \alpha({\text{max}(0,\frac{\text{mass}}{\text{masscon}} - 1.0)}^2) + \alpha({\text{max}(0,\frac{\text{cost}}{\text{costcon}} - 1.0)}^2)\\
& + \alpha({\text{max}(0,\frac{\text{time}}{\text{timecon}} - 1.0)}^2) + \beta(\text{milling loss})^2 + \lambda(\text{milling loss})   
\end{split}
\label{eq:sub3obj}
\end{equation}\\
where the notation definitions for $c$, $c_0$, mass, masscon, cost, costcon, time, timecon are given in Table \ref{tab:obj} and $\alpha$ is the penalty coefficient, $\beta$ is the milling loss penalty coefficient and $\lambda$ is the Lagrange multiplier.
We find that utilizing the concept of the Augmented Lagrangian method for the 3-axis milling constraint violation gives better results compared to just a penalty method. In each iteration, the Lagrange multiplier $\lambda$ is updated as follows:\\
\begin{equation}
    \lambda = \lambda + \gamma(\text{milling loss})
\end{equation}\\
where we find empirically that the best results are obtained when $\gamma$ follows the schedule: Initialize $\gamma = 0$. Increment by 0.1 in each optimization iteration until $\gamma = 10$, and keep $\gamma = 10$ for the remaining iterations.

We use the concept proposed in \cite{langelaar2019topology} for topology optimization with milling constraints. For a part to be manufactured by 3-axis milling, all regions of this part should be able to be reached by the milling tool, and thus, there should not be any void regions present in the part. We calculate the milling loss as shown in Algorithm \ref{alg:3axisloss}.
\begin{algorithm}
\caption{3-axis milling loss}\label{alg:3axisloss}
\begin{algorithmic}[1]
\State Initialize number of milling directions: $n$ \Comment{$n \leq 6$}
\State Initialize list of milling directions: $md$ \Comment{If $n = 6$, then $md = [x+,x-,y+,y-,z+,z-]$}
\State Initialize $i = 0$
\State Initialize $\text{xPhys} = \textit{Den}(\mathbf{X}_{den})$\Comment{xPhys is the tensor representing the density values at each element of the discretized part at the current iteration of topology optimization}
\State Initialize $\text{loss} = \mathbf{1}$ \Comment{loss is a tensor of ones having the same shape as the discretized part}
\Function{MLOSS}{axis}
\State $cs \leftarrow \text{CumulativeSummation}(\text{xPhys}, \text{axis})$ \Comment{Use cumulative summation along the given axis to find a region that is not accessible by the tool. Once a part surface starts, i.e., the density of some element is close to 1, all the remaining elements in the axis direction now have values close to and above 1.}
\State $hs \leftarrow \frac{1}{(1+\exp^{-10(cs - 0.5)})}$ \Comment{Use a Heavyside function to ensure all density values are between 0 and 1}
\State $\text{loss} = hs \odot (1-\text{xPhys})$ \Comment{The loss (penalty) is only for regions that are within the part and not the surrounding regions. $\odot$ is Hadamard product}
\State \Return loss
\EndFunction
\For {$n$ iterations}
\State $\text{axis} \leftarrow md[i]$ \Comment{$md[i]$ corresponds to the $i^{th}$ milling direction in $md$}
\State $\text{loss} = \text{loss}\odot\text{MLOSS(axis)}$ \Comment{Calculate the intersection of loss tensors for each milling direction}
\State $i \leftarrow i + 1$
\EndFor
\State 3-axis milling loss $= \text{Mean}(\text{loss})$ \Comment{Find the mean (sum over values of all elements in loss tensor and divide by number of elements) to get the final loss value}
\end{algorithmic}
\end{algorithm}

The compliance $c$  and mass are calculated in the same way as in Additive Manufacturing.

For the nominal time ($t_{nM}$), we use the following equation:\\
\begin{equation}
    t_{nM} = t_{sM} + t_{fM} + t_{mM} + t_{pM} + t_{iM}
     \label{eq:subtime}
\end{equation}
where $t_{sM}$ is the machine setup time, $t_{fM}$ is the fixture setup time, $t_{mM}$ is the machining time, $t_{pM}$ is the polishing time and $t_{iM}$ is the inspection time.
The machining time is calculated as follows:\\
\begin{equation}
    t_{mM} = \frac{V_r}{Q_v}
\end{equation}
where $V_r$ is the machined volume and $Q_v$ is the volume based removal rate.

All other terms in the Equation \ref{eq:subtime} are considered as constants (we use Haas DM-3axis as a reference) in the examples shown in section \ref{Results} and can be changed by the user if required.

The nominal cost ($c_{nM}$) is calculated as follows:
\begin{equation}
    c_{nM} = c_{sM} + c_{fM} + t_{mM}\times c_{mM} + c_{pM} + c_{iM} + c_{matM}
    \label{eq:subcost}
\end{equation}
where $c_{sM}$ is the machine setup cost, $c_{fM}$ is the fixture setup cost, $t_{mM}$ is the machining time (in minutes), $c_{mM}$ is the machining cost per minute, $c_{pM}$ is the polishing cost, $c_{iM}$ is the inspection cost and $c_{matM}$ is the material cost.

The machining cost per minute is a standard value we input. The material cost is the mass of the block to be machined (in kg) times the material cost per kg. All other terms in the equation are considered constants (Haas DM-3axis is used as a reference).

We perform probing similar to as shown in additive manufacturing. The cost and time for 3-axis milling are negatively correlated to the volume fraction of a part because more machining must be done to achieve a lower volume fraction part. Hence, unlike in additive manufacturing, where there was a positive correlation, the volume fraction corresponding to the active constraint is now calculated differently. If the volume fraction corresponding to the mass constraint ($vf_{mass}$) is the highest, then the mass constraint is the active constraint, and we can use $vf_{mass}$ in calculating $\mathbf{o_1}$ and $c_0$. If the volume fraction corresponding to the actual cost or lead time constraints is higher than $vf_{mass}$, then it indicates that the optimization is infeasible with the given constraints. This is because the above scenario implies that for achieving $vf_{mass}$, the cost or time required is more than the given cost or time constraint. Hence, we can eliminate this scenario for design optimization and save time and computational resources. In the examples shown in section \ref{Results}, we eliminate the options with infeasible constraints using the above logic. We can also use some error percentage $e_p$ for elimination, wherein even if the volume fractions corresponding to the actual cost and lead time are $e_p$ greater than $vf_{mass}$, we proceed with the design optimization to avoid early elimination.

We use the same schedule for penalty coefficient $\alpha$ as used in the Additive Manufacturing module.\\
\textbf{2-axis cutting:}~\\
The manufacturing process considered is as follows:\\
$\text{Machine setup} \rightarrow \text{Cutting Operation} \rightarrow \text{Polishing} \rightarrow \text{Inspection}$

We consider EDM (Electrical Discharge Machining) as the 2-axis cutting process for our model. The objective function is the same as Equation \ref{eq:addobj}, with the cost and time defined differently as follows:

For the nominal time ($t_{nEDM}$), we use the following equation:
\begin{equation}
    t_{nEDM} = t_{sEDM} + t_{cEDM} + t_{pEDM} + t_{iEDM}
    \label{eq:2axistime}
\end{equation}
where $t_{sEDM}$ is the machine setup time, $t_{cEDM}$ is the cutting time, $t_{pEDM}$ is the polishing time and $t_{iEDM}$ is the inspection time.
The cutting time is defined as follows:
\begin{equation}
    t_{cEDM} = \frac{A_{EDM}}{Q_{EDM}}
\end{equation}
where $A_{EDM}$ is the cutting area and $Q_{EDM}$ is the EDM feed rate.

A differentiable representation of the EDM cutting area is needed to incorporate it into the loss function. The density gradient $\frac{\partial\rho}{\partial X_{den}}$ of the topology can be calculated via automatic differentiation of the neural network, as shown in \cite{chen2023concurrent}. We filter this density gradient by using a Heavyside function $H_a$, which we find empirically to perform the best, to obtain magnitudes of 1 where the surface is present and 0 elsewhere. 

\begin{equation}
    H_a(x) = \frac{1}{1+e^{-x+5}}
\end{equation}

Then we obtain the total area for cutting as the summation of the output of the heavyside function for each element, over all the elements in the design domain.

\begin{equation}
    A_{EDM} = \sum H_a(|\frac{\partial\rho}{\partial X_{den}}|)
\end{equation}

We assume the EDM machine GF Machining Solutions AC Progress VP3 with a feed rate ($Q_{EDM}$) of $40 in^2/hr$ in the examples in section \ref{Results}.
For the nominal cost ($c_{nEDM}$), we use the following equation:
\begin{equation}
    c_{nEDM} = c_{sEDM} + t_{cEDM} \times c_{cEDM} + c_{pEDM} + c_{iEDM} + c_{mEDM} 
    \label{eq:2axiscost}
\end{equation}
where $c_{sEDM}$ is the machine setup cost, $t_{cEDM}$ is the cutting time (in minutes), $c_{cEDM}$ is the cutting cost per minute, $c_{pEDM}$ is the polishing cost, $c_{iEDM}$ is the inspection cost and $c_{mEDM}$ is the material cost.
The cutting cost per minute is a standard value we input. The material cost is the mass of the block (in kg) to be cut times the material cost per kg. For the other terms in the equation, we use constants based on GF Machining Solutions AC Progress VP3 as the reference EDM machine.

We currently use a similar probing procedure as in 3-axis milling. Note that for 2-axis cutting, the actual cost and lead time are not highly correlated to the volume fraction but rather to the area to be cut. We use the equation $(1 - vf)\times\text{total volume}$ for approximating the cutting area for a given probing volume fraction $vf$. This results in a very conservative estimate and eliminates design generation even with some feasible constraints. Creating a better probing model for 2-axis cutting is left for future work. 

\subsection{Explainable AI and Results Interface}

A key component of our approach is the ability to assist a designer in understanding the design
space of feasible alternatives, including the identification of key variables, correlations, and anti-
correlations, thresholds, and tradeoffs. This is facilitated through our design space visualization
tools that help “explain” why certain designs are determined to be optimal and how the outcome
of our design generation tools depends on the tradeoffs made across multiple dimensions of
concern.

\begin{figure*}
\centering

\includegraphics[trim={0.2cm 2.5cm 1cm 4cm},clip, width=\textwidth]{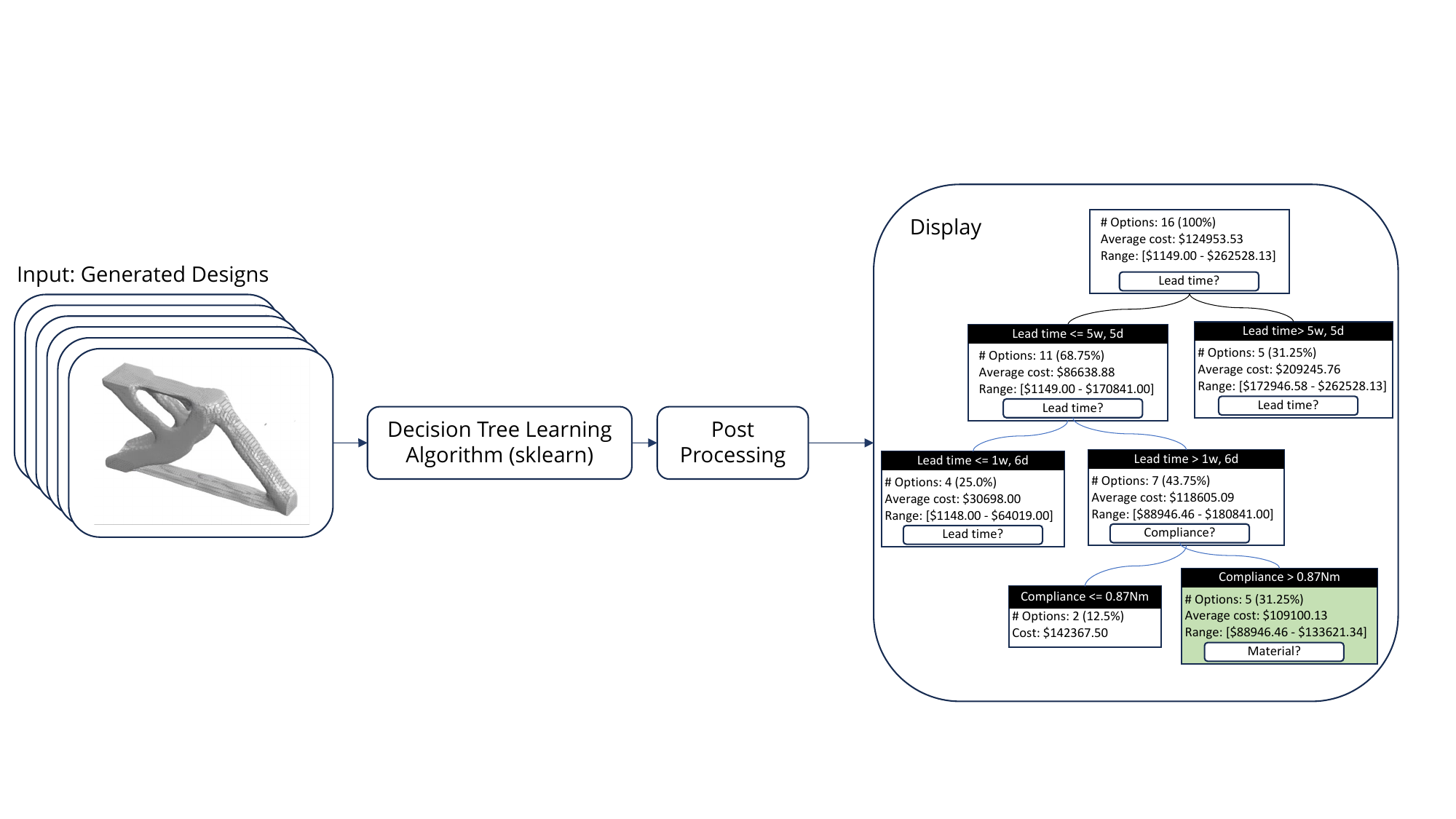}

\caption{The decision tree explainer. 1) All of the designs generated so far are used as inputs into a decision tree learner, implemented with the sklearn library. 2) The output from this is post-processed to provide a cleaner display of the decision logic and relevant information about the design space covered by each subtree. Subtrees containing designs from the current iteration are highlighted in green.}

\label{fig:dt}
\end{figure*}

The primary visualization is constructed using decision trees, which divide the design space into important decision points that provide natural partitions in the design. Such learned decision trees~\cite{Breiman2017},
 can be used to explain how a particular quality is impacted by the other qualities of
interest. Figure~\ref{fig:dt} shows the process we use to produce the decision trees, with an abbreviated decision tree on the right. 

The decision tree provides a combined view
about how a particular quality (in this case cost) is impacted by all the other concerns. This
allows a designer to understand what parts of the design space might be missing from
consideration to generate options in the next iteration, or to understand how many similar
options may exist in a particular part of the design space, as well as indicating thresholds (for
numerical variables) and decisions (for categorical variables) that influence the cost of design.
Alternative decision trees can be created to ``explain'' other variables, such as how the time
required for manufacture is affected by other variables, such as the choice of material or
compliance.

\section{Results}\label{sec4}
\begin{figure*}
\centering

\includegraphics[width=\textwidth]{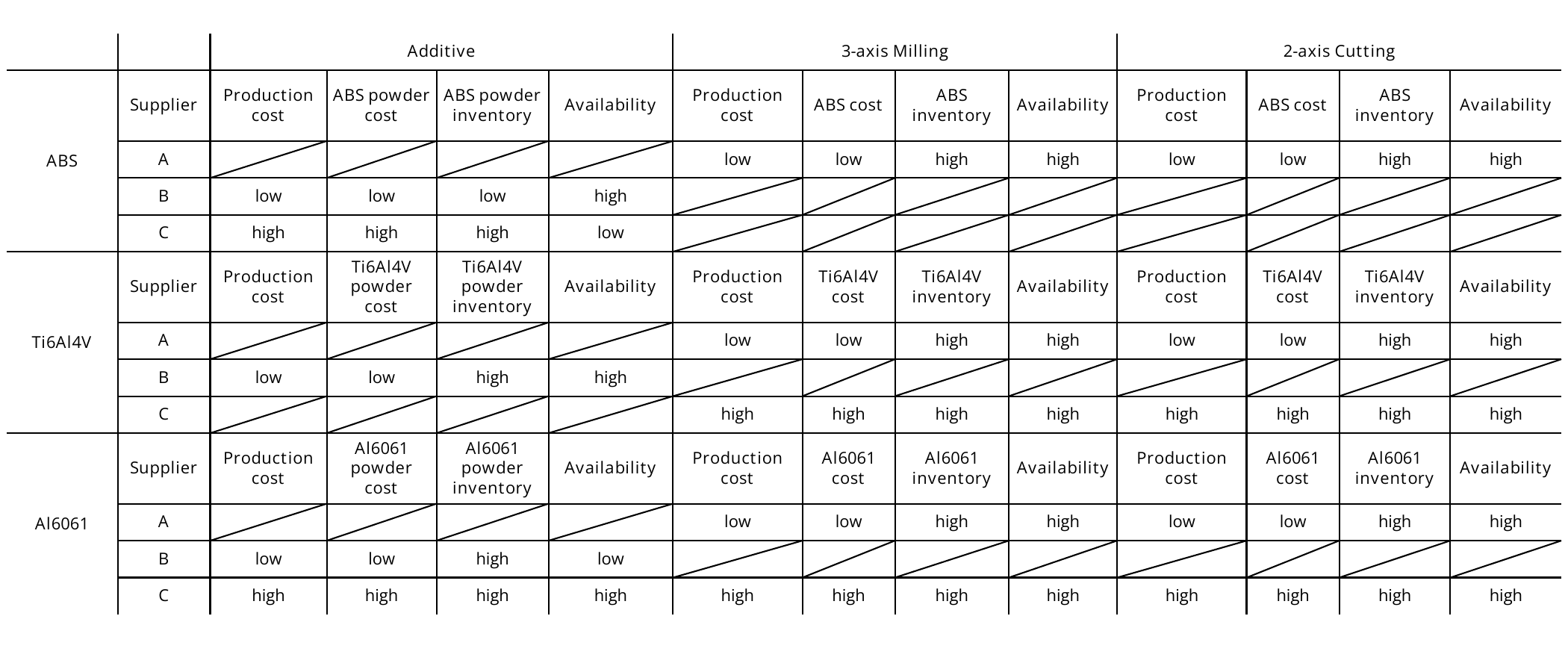}

\caption{Supplier configuration for the cantilever beam bracket and rocket engine mount studies.}

\label{fig:supplier_model}
\end{figure*}

\label{Results}
To demonstrate the application and performance of the GM framework, we set up two test cases. We assigned the dimensions of the two test cases to resemble medium and large size parts. The manufacturing methods available are 3-axis milling, additive manufacturing, and 2-axis cutting. Three total suppliers, Suppliers A, B, and C, are configured to represent the available machining facilities. To simplify our analysis of generative manufacturing, we restrict attention to part quantities that can be handled by single supplier bids. For enhanced clarity, we configured Supplier-A to be 3-axis milling and 2-axis cutting, Supplier-B to be additive only, and Supplier-C to provide both options. Cost and time factors are assigned to each of the suppliers. A detailed breakdown of the capabilities of the three suppliers is illustrated in Figure \ref{fig:supplier_model}. For brevity, in this section, we use the term `time' to indicate the total lead time and `cost' to indicate the total cost. In the previous section, we use EDM as an instance of 2-axis cutting. The material in the case studies includes ABS plastics which are not conductive; therefore, EDM machining cannot be performed. However, other 2-axis cutting processes, for example, water jet can be used as a substitute for which we use the same cutting rate and cost model. In this work, we limit the part orientation along the principal axis (x,y,z). This means that a total of six print and 3-axis milling orientations are possible: (x+,x-,y+,y-,z+,z-). For 2-axis cutting, the orientation of (x,y,z) is available. 

\subsection{Cantilever Beam Bracket}
\begin{figure*}
\centering

\includegraphics[width=\textwidth]{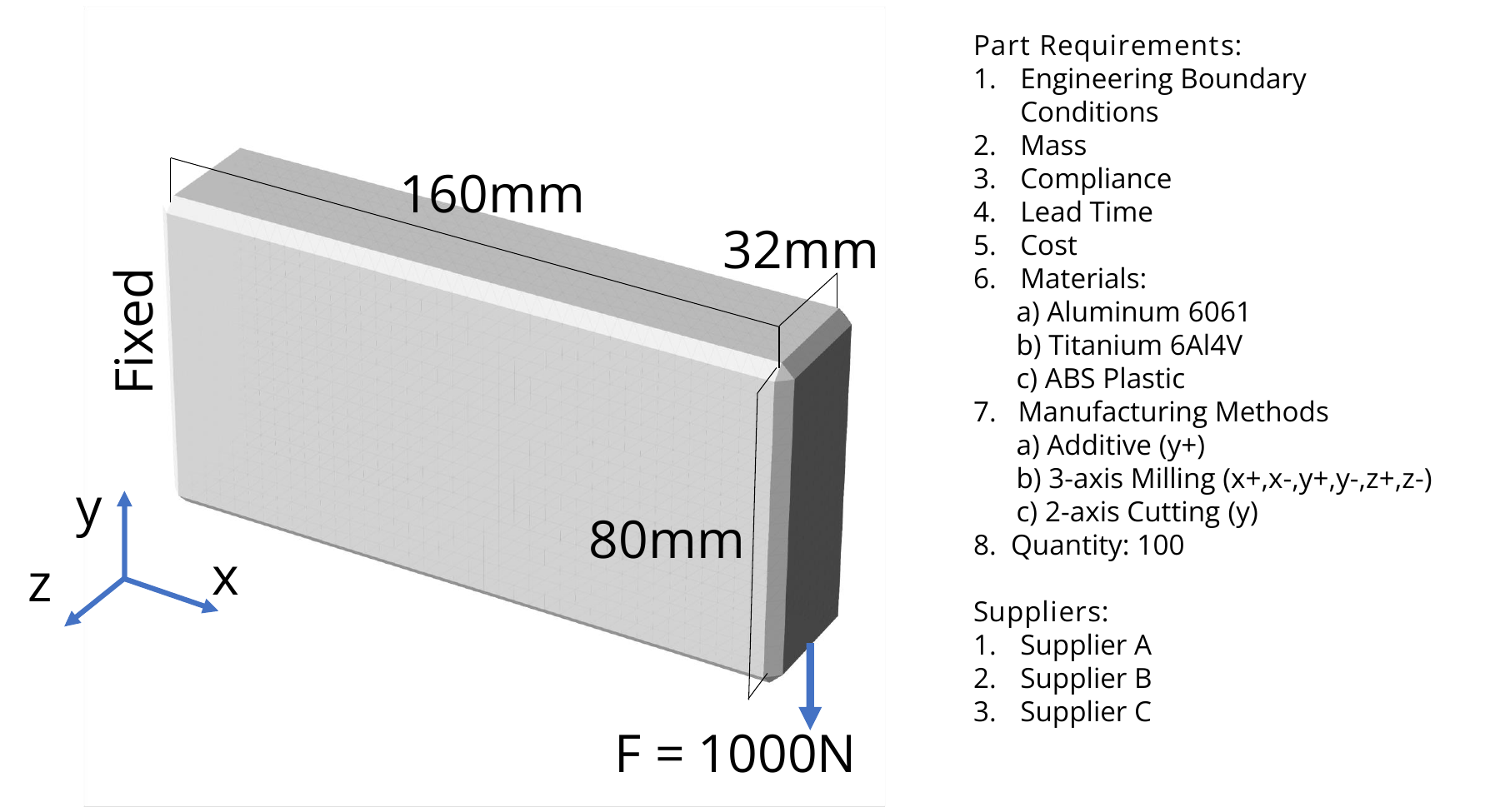}

\caption{The boundary conditions, dimensions, and the part requirements for the bracket example. }
\label{fig:bracket_bc}
\end{figure*}

 The cantilever beam is often seen in topology optimization papers as an example to compare the structural performance of the designed part. The problem's dimension and boundary conditions of the problem are illustrated in Figure \ref{fig:bracket_bc}. We envision a typical use-case for this type of bracket is medium-sized structural components with a total production of around 100 for each request to the supplier. For manufacturing orientation, we consider $y+$ for additive, all six orientations for 3-axis milling, and $y$ for 2-axis cutting.  

Once the specification is made, we can probe the supplier by creating a surrogate process plan and send it to the suppliers. These requests can be generated based on the selected manufacturing method, material, and structural performance requirement by the engineer. The supplier will respond to the request with a bid. The supplier response can give the engineer a relatively fast response with knowledge of the current manufacturing availability, which may help the engineer to refine the constraint specification further before running any relatively computationally expensive generative design. Furthermore, this helps the design optimizer by eliminating options that are not feasible to manufacture such that a subset of the total possible material, supplier, and manufacturing method combinations is being optimized.

The probing result is summarized in Table \ref{tab:bracket_itr1}. From the probing alone, we can identify that due to machine availability, none of the 2-axis cutting options are available from the suppliers. The Ti6Al4V subtractive options cannot be realized due to the low mass constraint. 
\begin{table}
    \centering
    \begin{tabular}{cccc}
         &  Additive (y+)&  3-axis Milling(x+,x-,y+y-,z+,z-)& 2-axis Cutting (y)\\
         Al6061&  \cmark &  \cmark& \xmark\\
         Ti6Al4V&  \cmark&  \xmark& \xmark\\
         ABS Plastic&  \cmark &  \cmark& \xmark\\
    \end{tabular}
    \caption{Probing supplier result. For Ti6Al4V 3-axis Milling (x+,x-,y+,y-,z+,z-): the mass constraint is very low, such that to buy and machine so much mass, the minimum cost is much greater than the cost constraint for all suppliers. Based on the supplier configuration, none of the 2-axis cutting options is available.}
    \label{tab:bracket_itr1}
\end{table}

\begin{figure*}
\centering

\includegraphics[width=\textwidth]{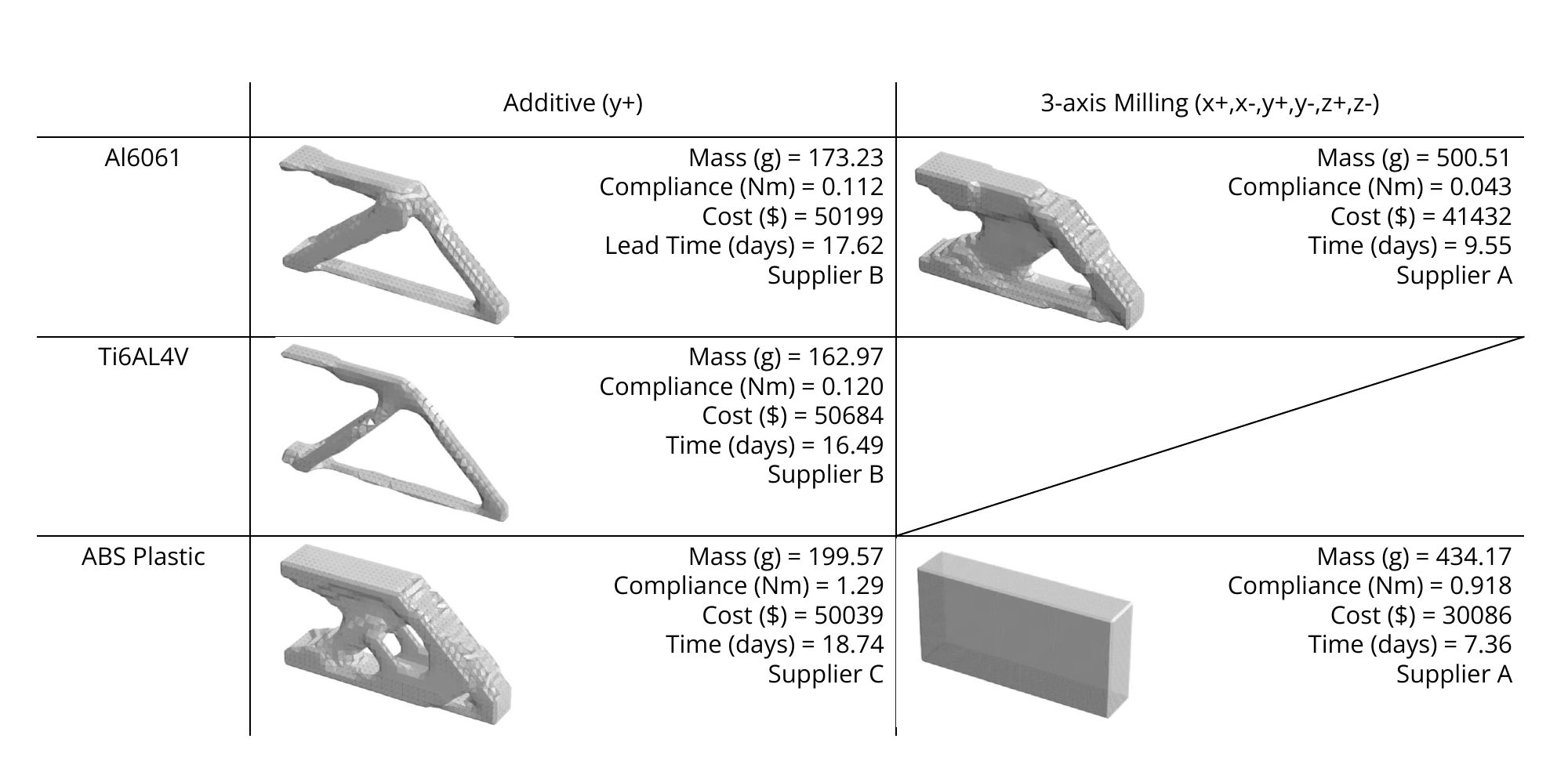}

\caption{For the bracket example, a total of 100 parts are ordered. We set up the constraints as follows: the mass should be smaller than 500~g, the total cost should be less than \$50000, and time should be less than 1 month. In the table, we include all the objective values achieved for the best supplier.  }

\label{fig:bracket_case}
\end{figure*}

Suppose the engineer is satisfied with the probing results and no further adjustment to the constraint is desired. In that case, the generative design can be performed for all available materials and manufacturing combinations. We summarize the result of the study in Figure \ref{fig:bracket_case}. The result demonstrated variation across manufacturing methods and materials. For the given set of constraints and the objective of maximizing stiffness, all additive manufacturing solutions have the cost constraint as the active constraint, indicating loosening the cost constraint can give stiffer solutions. For 3-axis milling manufacturing solutions, the mass constraint is the active constraint here. Removing more material from a solid block requires more cost and time and results in a part that is less stiff. Our system shows that for this bracket example and supply chain situation considered, for ABS Plastic (which is lighter compared to Al6061 and Ti6AL4V), for 3-axis milling, a solid block, which is the starting point of the milling operation, satisfies all the given constraints. Hence, our system rightly presents the optimal solution as the starting block itself. Each of the solutions presented is of the best supplier. For example, Supplier B gave the best solution in terms of objective value and constraint satisfaction, and hence, we show Supplier B's solution. We present a detailed analysis of solutions obtained for different suppliers and factors in choosing one over the other in section \ref{sec:remeg}, Figure \ref{fig:engine_itr1_suppliers}.

Finally, based on all the solutions presented, the engineer will either decide on which solution to pursue or use the result to guide refinement on the constraint value selection to further narrow down the candidates. In the rocket engine mount case study, we will explore the iterative refinement of the constraint value. 


\subsection{Rocket Engine Mount}\label{sec:remeg}
\begin{figure*}
\centering

\begin{subfigure}[t]{0.3\textwidth}
\centering
\includegraphics[width=\textwidth]{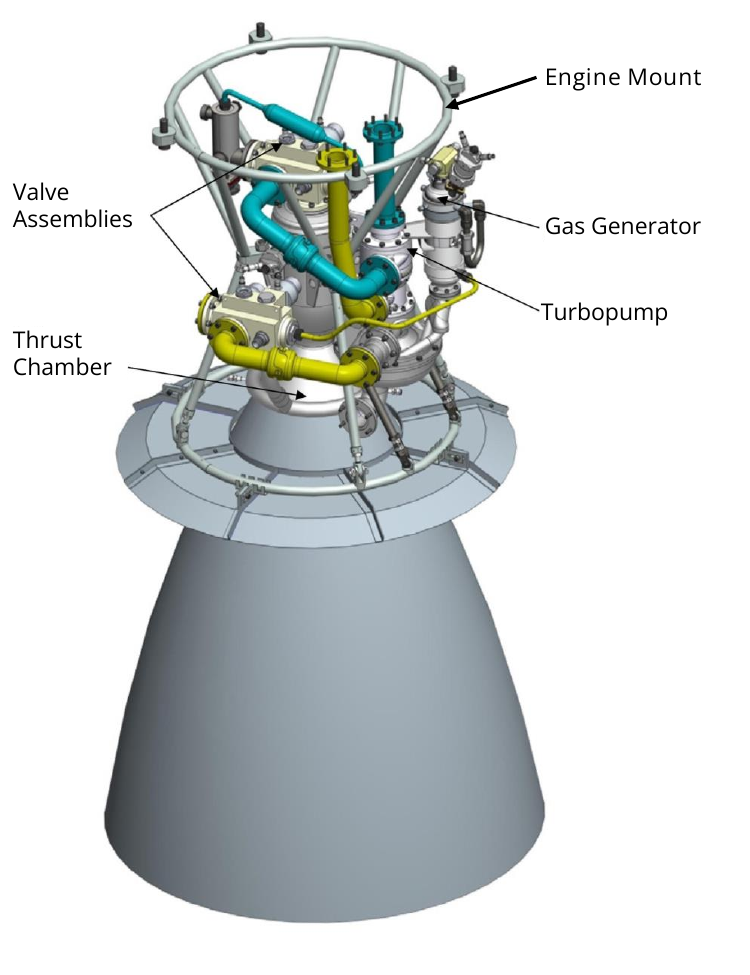}
\end{subfigure}
\qquad
\begin{subfigure}[t]{0.6\textwidth}
\centering
\includegraphics[width=\textwidth]{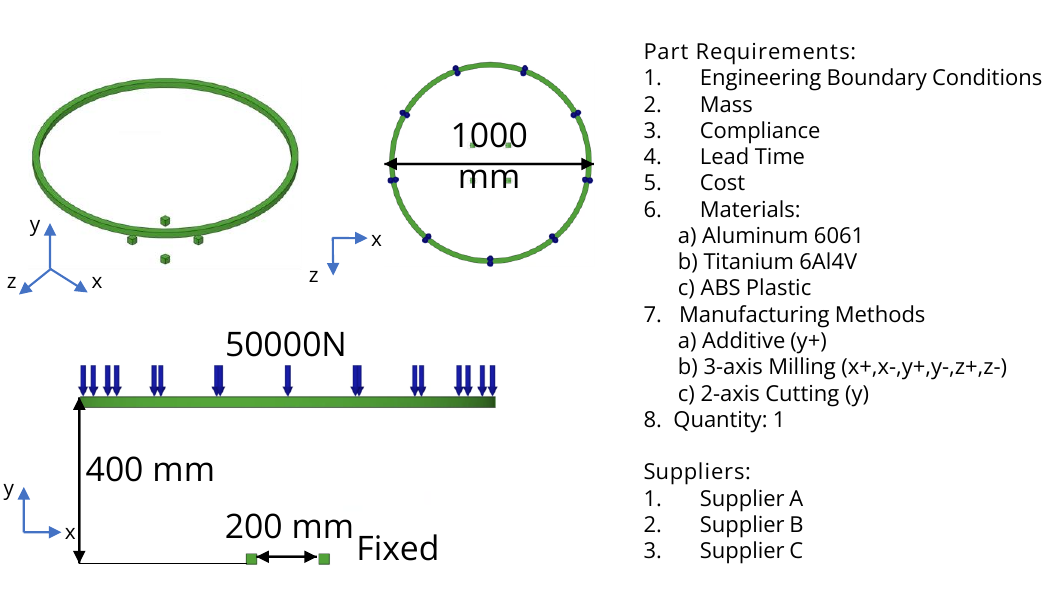}
\end{subfigure}

\caption{The boundary conditions and the dimensions for the rocket engine mount example. The engine rendering is based on the work by Almeida and Pagliuco \cite{almeida2014development} where the engine mount is located on top of the assembly. }

\label{fig:engine_bc}
\end{figure*}

The second example is a rocket engine mount. The mount is configured to transfer the thrust from the engine to the fuel tanks. We are inspired by the work \cite{almeida2014development} where the rocket engine consists of four mounting holes. We reduce the size of the engine and tank so that the engine mount can be subtractively machined as a single piece without assembly. The diameter of the tank is 1~m, and the mounting holes are spaced 20~cm apart. We assume the engine is outputting a thrust of 50~kN. The boundary condition is configured such that the four mounting holes are fixed. The resulting reaction from the tank is modeled as a 50~kN force applied on a thin ring on top. The boundary condition and the dimension of the engine mount are illustrated in Figure \ref{fig:engine_bc}. Based on the geometry of the engine mount, we identify the manufacturing orientation for the three manufacturing methods. In additive, we choose the (y+) direction as the print orientation where the part is the lowest in height. In 3-axis milling, due to the potentially complex geometry generated, all six orientations are selected. For 2-axis cutting, we select the cutting direction to be (y). 

Similar to the bracket example, the engineer first starts by defining the constraints and manufacturing methods. Then, the suppliers are selected. The probing result for the first iteration is summarized in Table \ref{tab:engine_itr1}. From the probing result, we can observe that the 3-axis milling and 2-axis cutting manufacturing methods for Ti6Al4V are not feasible due to the heavy stock material required to purchase.

\begin{table}
    \centering
    \begin{tabular}{cccc}
         &  Additive (y+)&  3-axis Milling (x+,x-,y+y-,z+,z-)& 2-axis Cutting (y)\\
         Al6061&  \cmark &  \cmark& \cmark\\
         Ti6Al4V&  \cmark&  \xmark& \xmark\\
         ABS Plastic&  \cmark &  \cmark& \cmark\\
    \end{tabular}
    \caption{Probing supplier results for iteration 1. For Ti6Al4V 3-axis milling: the mass constraint is very low, such that to buy and machine so much mass, the minimum cost is much greater than the cost constraint, for all suppliers. Hence, this configuration is deemed infeasible, and further calculation can be avoided}
    \label{tab:engine_itr1}
\end{table}

\begin{figure*}
\centering

\includegraphics[width=\textwidth]{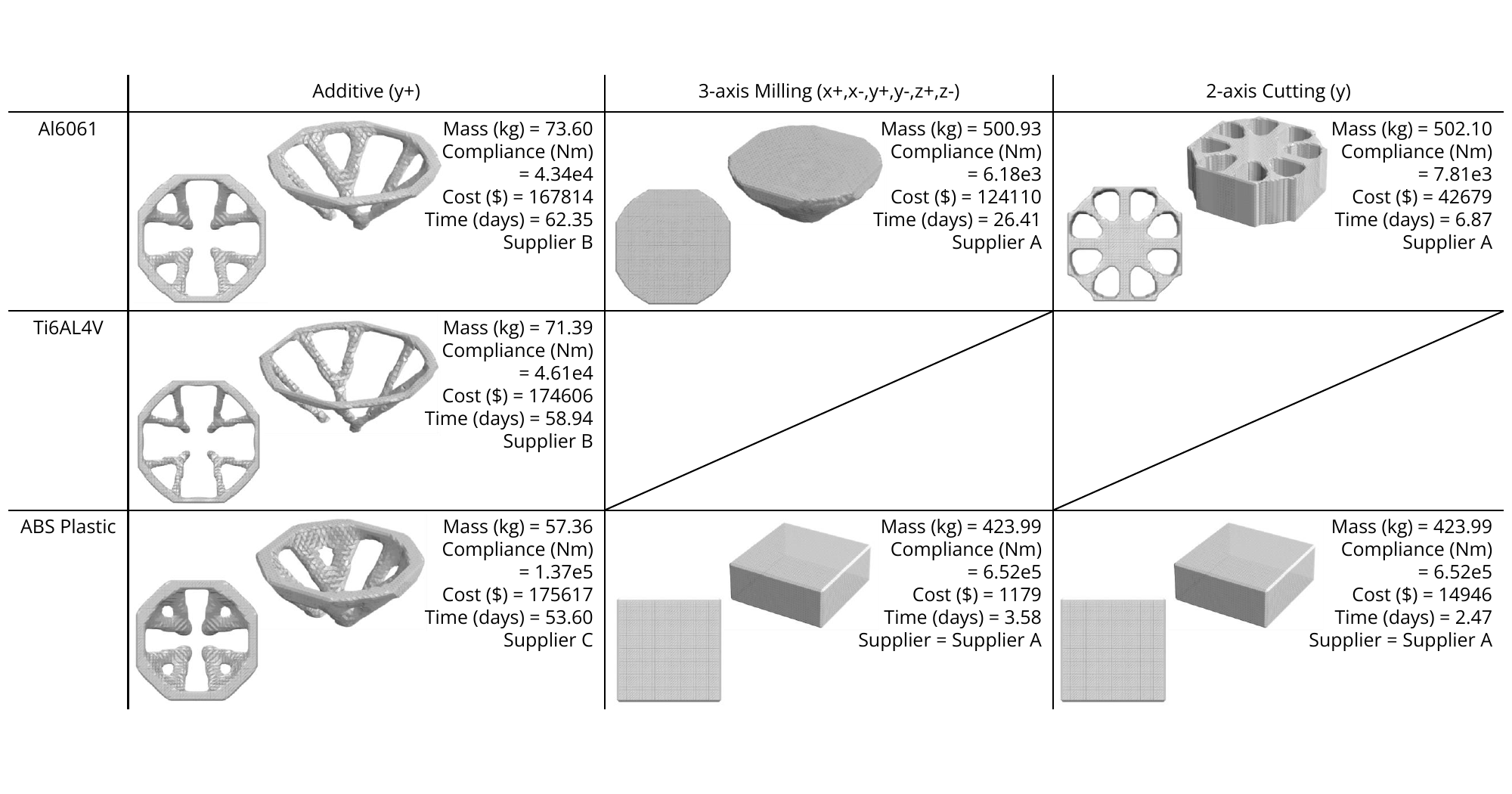}

\caption{For the rocket engine mount example, a single part is ordered. In the first iteration, we set up the constraints as follows: the mass should be smaller than 500~kg, the total cost should be less than \$175000, and the time should be less than 2 months (61~days). }

\label{fig:engine_itr1}
\end{figure*}

Next, the generative design optimization can commence. The result is summarized in Figure \ref{fig:engine_itr1}. As with the bracket, we can see a similar tendency between additive and subtractive processes. Given the objective of compliance minimization, the additive solutions reached the cost constraint and subtractive solutions reached the mass constraint. 

\begin{figure*}
\centering

\includegraphics[width=\textwidth]{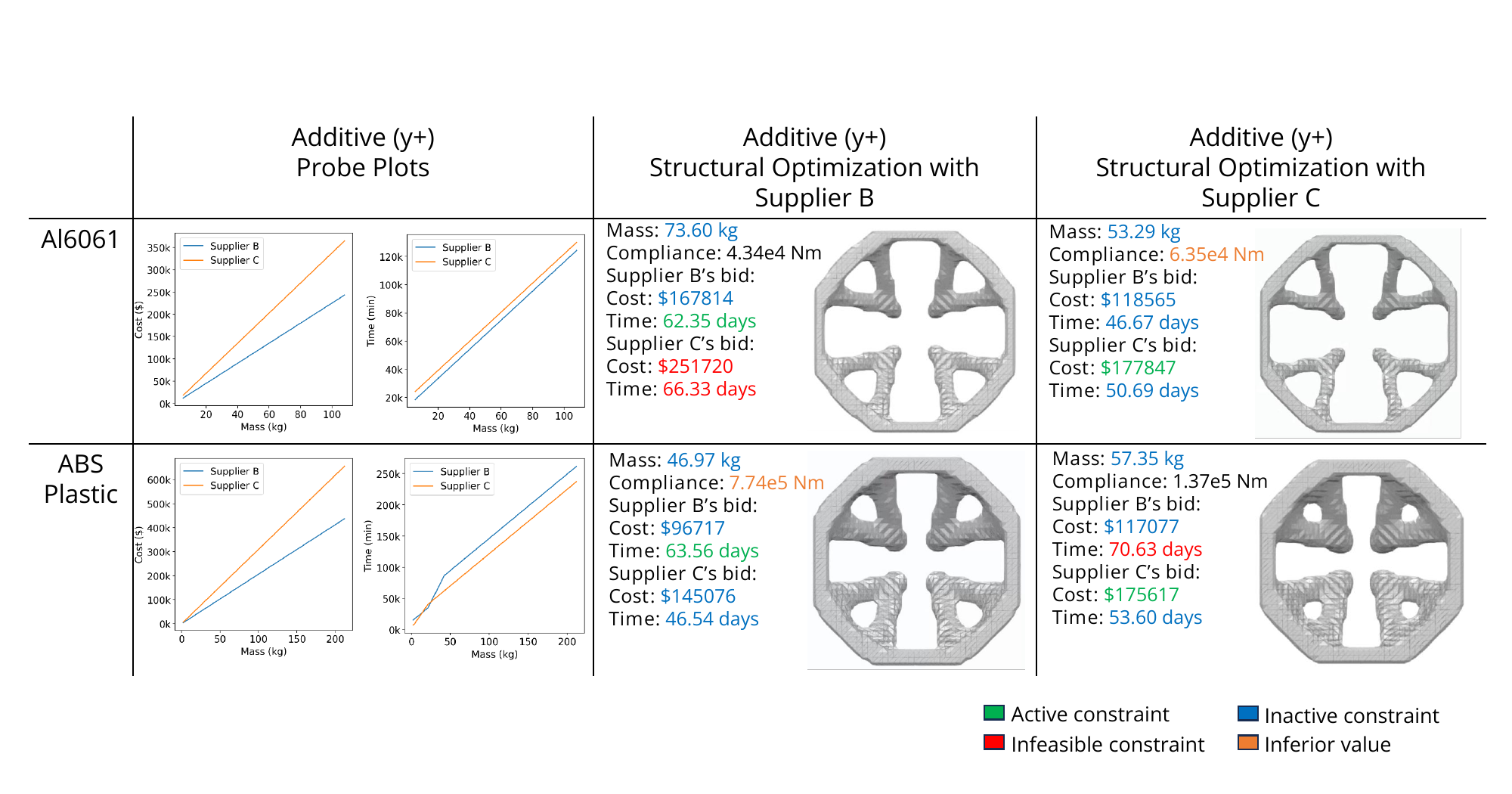}

\caption{We can also run the topology optimization for each supplier. For example, with additive manufacturing method and Al6061, even though the cost and time may be higher for supplier C, and thus the structure that satisfies the cost and time constraints is theoretically less rigid than what can be achieved with topology optimization considering Supplier B, Supplier C might be more trustworthy, or there may exist other factors that influence the decision towards Supplier C. The user can then evaluate using our system if these other factors have more weightage than the loss in theoretical structural rigidity.
}

\label{fig:engine_itr1_suppliers}
\end{figure*}

In Figure \ref{fig:engine_itr1}, we show the solution for the best supplier. For example, Supplier B is the best for Additive (y+) and Al6061 in terms of the objective value and constraint satisfaction and gives the solution presented in the corresponding cell in the figure. We can also run the topology optimization for each supplier, where the result is summarized in Figure \ref{fig:engine_itr1_suppliers}. For instance, using additive manufacturing and Al6061, even if Supplier C has higher costs and longer production times, resulting in a structure that is theoretically less rigid compared to what topology optimization with Supplier B can achieve, Supplier C might be more reliable, or there could be other considerations influencing the choice of Supplier C. The user can then use our system to assess whether these other factors outweigh the reduction in theoretical structural rigidity.

\begin{table}
    \centering
    \begin{tabular}{cccc}
         &  Additive (y+)&  3-axis Milling(x+,x-,y+y-,z+,z-)& 2-axis Cutting (y)\\
         Al6061&  \cmark &  \xmark& \xmark\\
         Ti6Al4V&  \xmark&  \xmark& \xmark\\
         ABS Plastic&  \cmark &  \xmark& \xmark\\
    \end{tabular}
    \caption{Probing supplier results for iteration 2 where mass and cost are further reduced. For Ti6Al4V Additive (y+), the cost of material is high, and it will not be possible to manufacture a part in the given cost constraint that is above a certain mass threshold for all suppliers. }
    \label{tab:engine_itr2}
\end{table}

Based on the result from iteration 1, we further reduce the mass and cost constraint and instigate another iteration. With the reduced constraint, probing results only show additive manufacturing as a viable option which is shown in Table \ref{tab:engine_itr2}.


\begin{figure*}
\centering

\includegraphics[width=\textwidth]{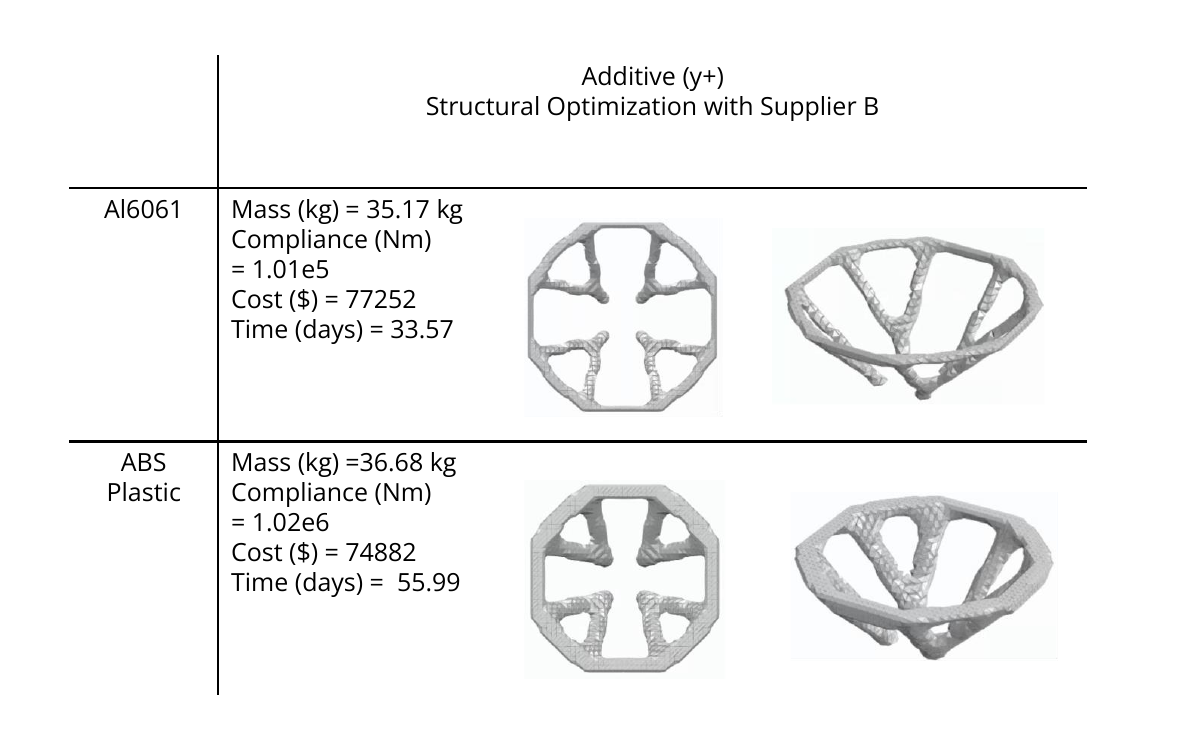}

\caption{Based on the result from the first iteration, we updated the constraint requirements to reduce the mass to 50~kg and a total cost of \$75000. Now only two combinations provided valid bids where the lower mass requirement benefited additive manufacturing.   }

\label{fig:engine_itr2}
\end{figure*}

The result from iteration 2 is summarized in Figure \ref{fig:engine_itr2}. Comparing the Al6061 and ABS plastic, Al6061 demonstrated superior structural performance with lower compliance. The slightly higher cost than the constraint is due to the penalty in the objective function and the linear surrogate model obtained from probing the supplier. The penalty value in equation \ref{eq:sub3obj} is a large number but not infinite. The objective function uses the surrogate model for optimization. However, once the final process plan is generated, the actual nominal time and cost are used to create a process plan for which the time and cost are quoted. 

\begin{figure*}
\centering

\includegraphics[trim={15cm 3cm 14.5cm 2.5cm},clip, width=0.75\textwidth]{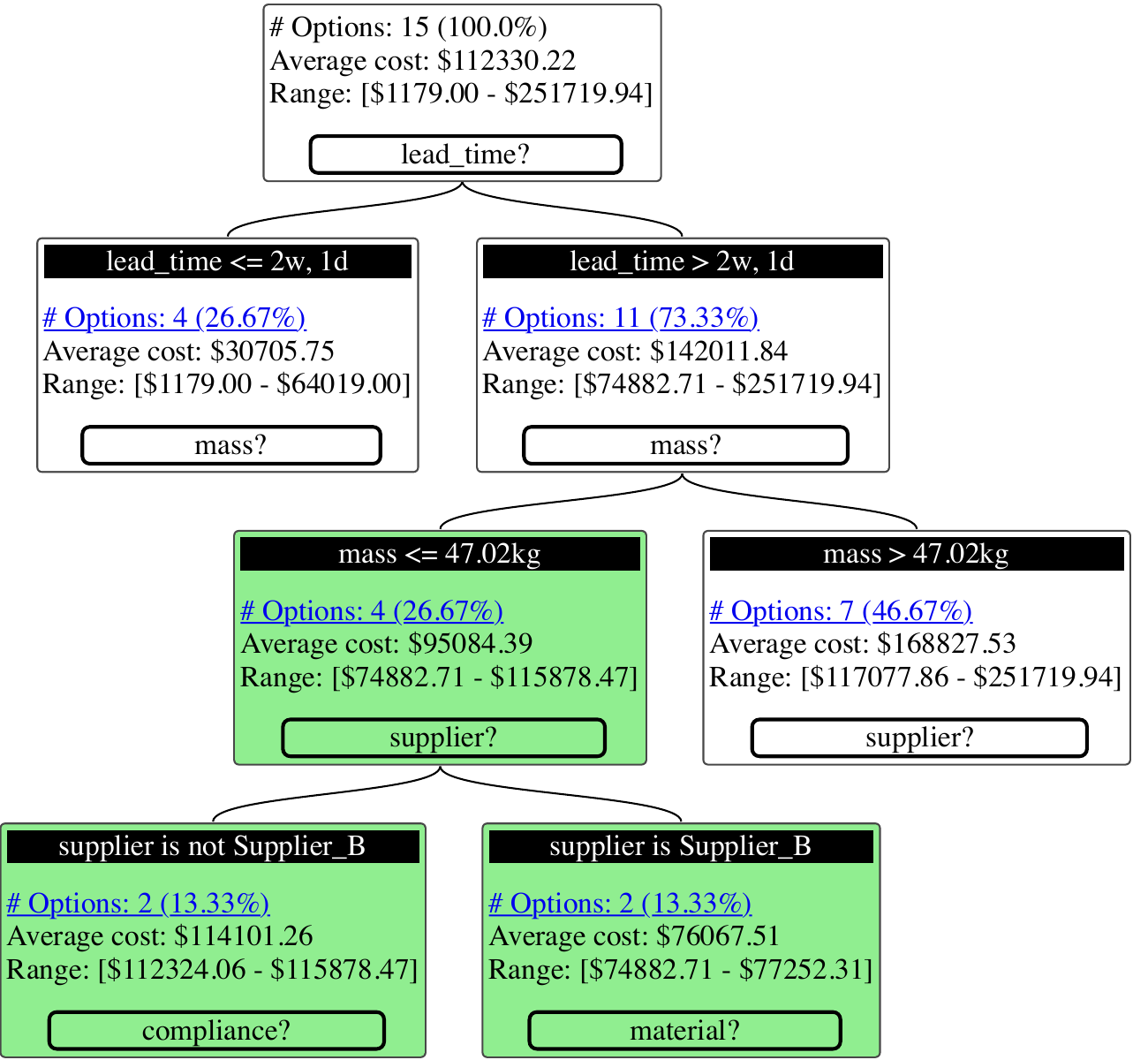}

\caption{The Decision Tree for the rocket mount example, showing how cost of manufacturing is related to decisions about lead time, compliance, and choice of material.}

\label{fig:dt_2}
\end{figure*}

In Figure~\ref{fig:dt_2}, we show the decision tree from this iteration with decisions related to cost. This particular decision tree informs the
designer about how the cost of manufacturing is impacted by decisions about lead-time, supplier, and
choice of manufacturing material. As illustrated, the top-level decision point is based on lead
time (i.e., the time to complete the manufacturing) -- meaning that decisions about lead time are
the most important with respect to differentiating cost, followed by mass, then whether the supplier is Supplier B. The decision tree also indicates how many of the
designs make similar decisions (which are those included in the same subtree), helping the designer understand clustering behavior. For example, following the tree to the left-most node (lead\_time $<=$ 2w, 1d), we can see that of the 15 options that have been generated by the two iterations, 26.67\% of them have a lead time of less than or equal to 2 weeks, 1 day. The designs that were generated in the 2nd iteration are contained in the green highlighted subtree, which can be examined further by clicking on the node to show them or by further expanding the subtree to show further details. We can see that the new iteration explored a part of the design space that was not in the first iteration and can easily see that none of these new designs have a lead time below 2 weeks and 1 day.

The final decision can be made if the engineer is satisfied with the result in iteration 2. Given the targeted application for a rocket engine mount, the engineer can choose the Al6061 version due to thermal requirements. The engine mount example demonstrates the versatility of our proposed GM framework in a case study. From the supply chain perspective, the information on the current material availability, time, cost, and scheduling of each supplier can be communicated with the design generator. Before running any design generator in each iteration, the surrogate model built from probing the supplier helps to eliminate options that cannot be achieved, while the correlation between time and cost versus nominal time and nominal cost helps the design generator create design candidates that satisfy the design constraints.

\begin{figure*}
\centering

\includegraphics[width=\textwidth]{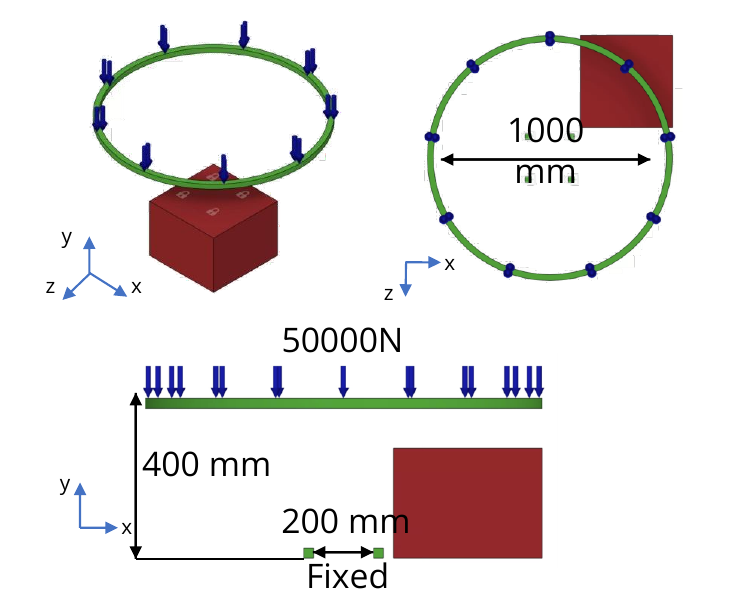}

\caption{Compared to the previous engine boundary condition, we added a no-design region on one quadrant. }

\label{fig:engine_bc1A}
\end{figure*}

\begin{figure*}
\centering

\includegraphics[width=\textwidth]{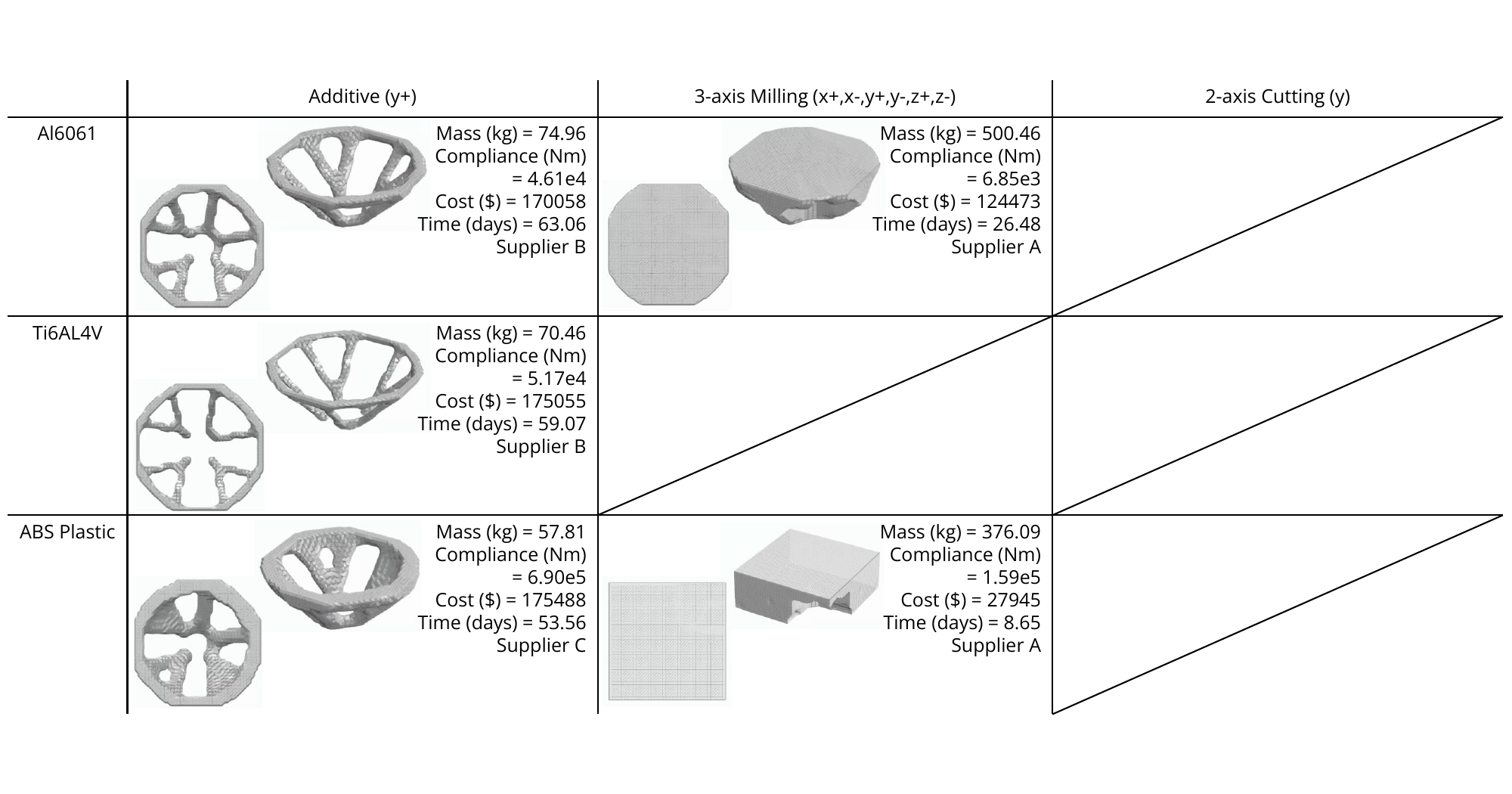}

\caption{In this example, the constraints remain the same as the first iteration of the previous case study. However, with the addition of the no-design region, 2-axis cutting is no longer feasible due to geometry constraints.  }

\label{fig:engine_case1A}
\end{figure*}

\subsection{Engine Mount Example with No Design Region}

No-design region adds additional design constraints to the optimization. It is often used when existing components intersect with the design domain. In this side study for the engine mount example, we prescribe a no-design region on one of the quadrants as illustrated in Figure \ref{fig:engine_bc1A}. All other constraints and requirements remain the same with iteration 1 of the previous example. 

The result from this case study is shown in Figure \ref{fig:engine_case2}. Due to the placement of the no-design region, none of the 2-axis cutting options is viable. We can also observe that due to the addition of no-design regions, other generated examples no longer demonstrate rotational symmetry.

\subsection{Engine Mount Example with Alternate Supplier Model}

A variety of factors may affect the performance of suppliers. Due to the addition of machine inventory or cancellation of a previous order, a supplier may see a sudden reduction in time and cost for the current order. These changes in the supplier landscape should in turn affect the result of generative manufacturing. In this example, we showcase how our system adapts to these changes and gives optimal solutions with respect to the requirements and resources. We change the supplier capabilities shown in Figure \ref{fig:supplier_model} such that now subtractive manufacturing is inexpensive and can be done at a quicker rate compared to additive manufacturing. We dramatically reduce the time and cost coefficients for subtractive manufacturing for Supplier A to simulate such an event. Furthermore, in this example, we tighten the constraints on time to 10~days, the mass to 75~kg, and the cost to less than $\$$25000. The probing results given the new supply chain situation and new constraints are given in Table \ref{tab:engine_case2}.

\begin{table}
    \centering
    \begin{tabular}{cccc}
         &  Additive (y+)&  3-axis Milling (x+,x-,y+y-,z+,z-)& 2-axis Cutting (y)\\
         Al6061&  \xmark &  \cmark& \xmark\\
         Ti6Al4V&  \xmark&  \xmark& \xmark\\
         ABS Plastic&  \xmark &  \cmark& \xmark\\
    \end{tabular}
    \caption{Probing supplier results for scenario 2 where subtractive manufacturing at Supplier A is inexpensive and with a minimal time. For Al6061 Additive (y+), the time is high, and it will not be possible to start manufacturing a part in the given constraint of 10 days, for all suppliers. }
    \label{tab:engine_case2}
\end{table}

\begin{figure*}
\centering

\includegraphics[width=\textwidth]{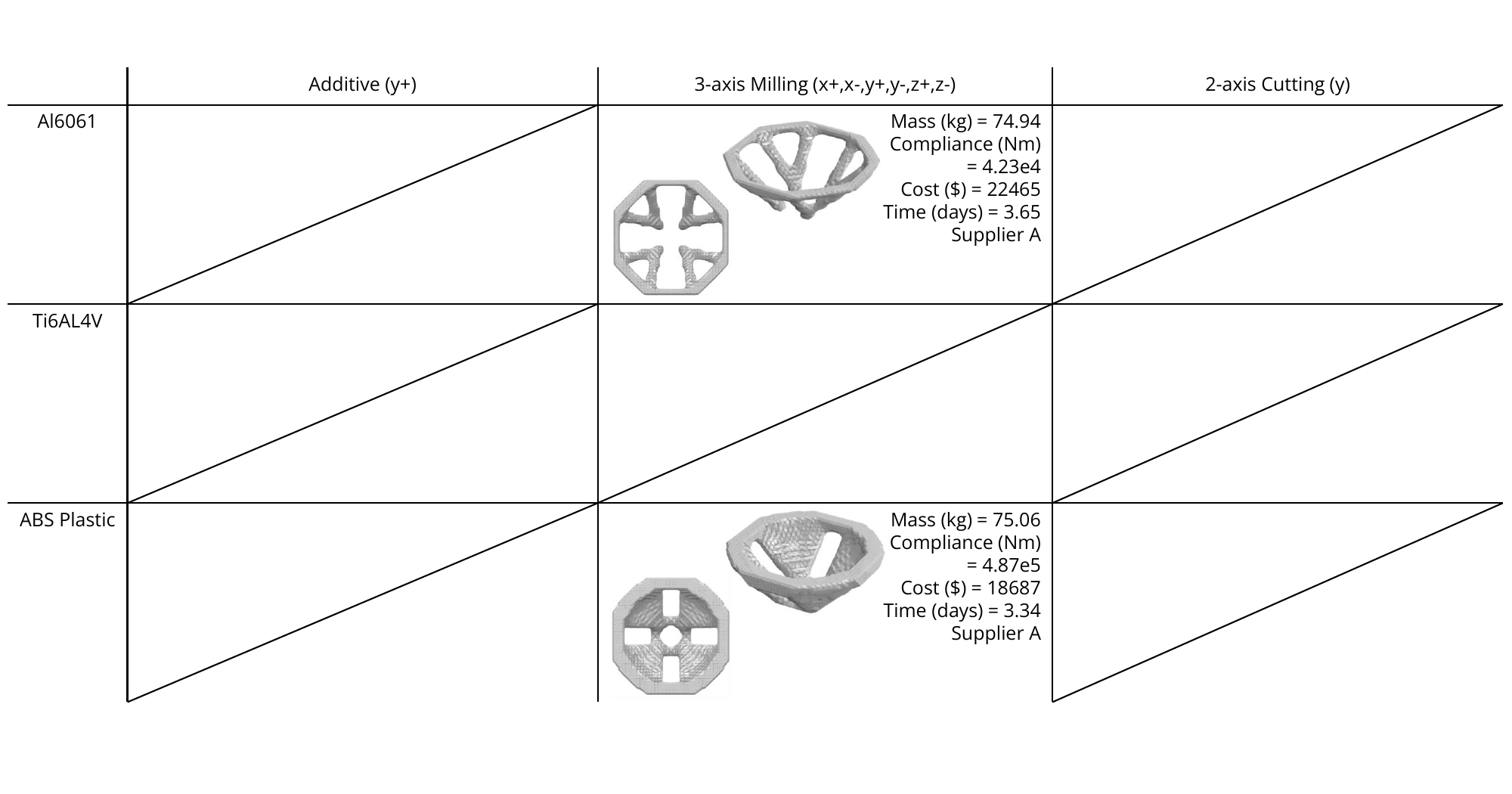}

\caption{If the supply chain situation changes, the optimal parts change, and we showcase our system’s results here. Subtractive Manufacturing at Supplier A has extremely low cost and lead time in this case study and the constraints are tighter: mass should be less than 75~kg, cost should be less than $\$$25000 and time should be less than 10 days, which make the other options infeasible.}

\label{fig:engine_case2}
\end{figure*}

From the probing result given the updated constraints, we can see that none of the additive options is feasible due to the short time requirement. However, due to the reduction in time and cost for Supplier A, bids from Supplier A became feasible. The optimization result is shown in Figure \ref{fig:engine_case2}. 

\begin{figure*}
\centering

\centering

\begin{subfigure}[t]{0.45\textwidth}
\centering
\includegraphics[width=\textwidth]{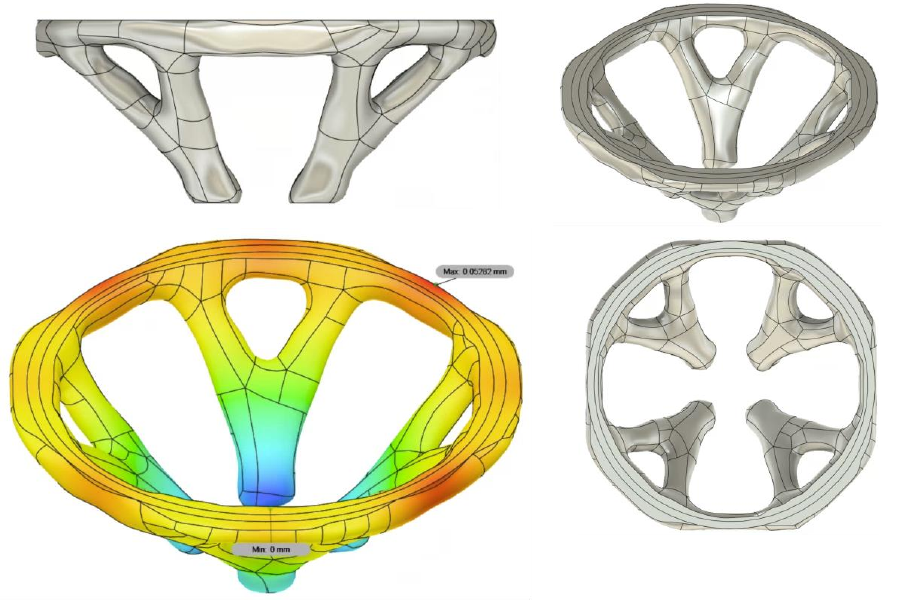}
\caption{Our method: m=98.35~kg, max displacement: 0.0528~mm}
\end{subfigure}
\qquad
\begin{subfigure}[t]{0.45\textwidth}
\centering
\includegraphics[width=\textwidth]{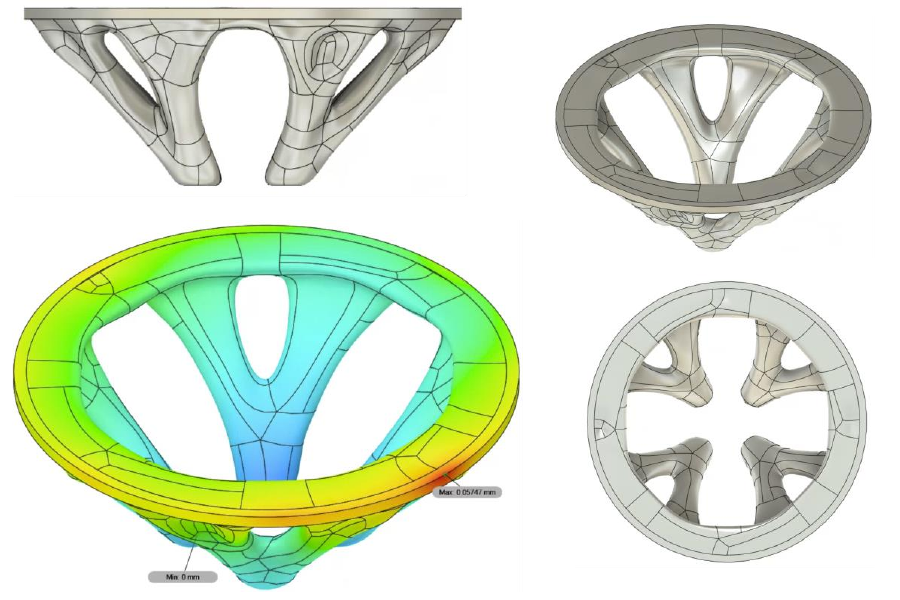}
\caption{Fusion 360: m=98.14~kg, max displacement: 0.0575~mm}
\end{subfigure}

\caption{Comparing the mass and maximum displacement between our method and Fusion 360 for topology optimization. The mass constraint is 100 kg and both methods satisfied the constraint, while the neural network-based topology optimization reached a slightly smaller displacement with a slightly higher mass. }

\label{fig:tocomp}
\end{figure*}
\subsection{Structural optimization benchmark}

The underlying neural network-based topology optimization allowed us to integrate design, manufacturing, and supply chain constraints. To verify the performance of the topology optimizer, we compare our implementation with commercial software Autodesk Fusion 360's generative design \cite{f360}. As each handles manufacturing constraints differently, we focus on comparing the topology optimization function alone without additional manufacturing constraints. The boundary condition is identical to the engine mount example as illustrated in Figure \ref{fig:engine_bc}. We evaluate the mechanical performance of both studies in Fusion 360 with their FE analysis tool. To export the neural network-based topology optimization result into Fusion 360, we first perform a marching cube analysis to extract the iso-surface of the geometry as a mesh. Then the mesh is imported to Fusion 360 and then converted to solid with t-spline analysis. Finally, the loading ring and the four bottom mounting points are added to the solid such that the load can be applied consistently in FE analysis. We observe that both methods are under the mass constraint of 100~kg while the neural network-based topology optimization reached a slightly smaller displacement with a slightly higher mass. This comparison demonstrates that the neural network-based topology optimization can reach a structural performance comparable to that of commercial software.

\section{Limitations and Future Work}\label{sec5}

While we demonstrated the capabilities of our generative manufacturing system on several examples to mimic the real-world process flow with different materials, a combination of suppliers, and a variety of manufacturing methods, there are still assumptions we made that may not completely reflect a typical design and manufacturing process. Smaller parts will likely fit inside the building envelope of the additive or subtractive machines; however, larger parts may call for segmentation and assembly. We do not consider assembly of parts and the corresponding optimization in our current system, and we leave that for future work. Though designs for 3-axis machining can be adapted to a 5-axis machining center, optimization for 3-axis machining is still more restrictive than the 5-axis. We plan to implement design optimization with 5-axis machining and tool size constraints in the future. Other potential areas of focus and future directions include improving the cost and time formulations, optimization of the orientation of the part in additive manufacturing, finding optimal setups in subtractive manufacturing and including thermal analysis and stress-constrained topology optimization in the generative manufacturing compiler. 

\section{Conclusion}\label{sec6}
Methods that inform and optimize the design based on the engineering as well as business requirements, such as the lead time and actual cost, have not seen much success. Simultaneously considering the design, manufacturing and supply chain requirements and resources is a difficult but crucial problem whose solution can be highly beneficial to numerous industries. We present the Generative Manufacturing compiler and showcase through various examples its capacity to produce optimal components by factoring in all the aforementioned considerations and constraints. We show how the best solution changes if the requirements change or the state of the suppliers changes and the trade-offs within the suppliers for a particular design. Our proposed compiler provides substantial benefits to a user performing the process of part making by enabling adaptation according to the situation and ensuring optimal solutions are generated.

\section{Acknowledgements}
This article has been approved for public release by Lockheed Martin PIRA CET2024070158. This work was supported by the Lockheed Martin Corporation (MRA19001RPS004). We would like to thank Bob Hermida and James L. Mathieson for useful discussions. We would also like to thank Javier C\'{a}mara, Rebekka Wohlrab, and Pakshal Shah for their help on the explainability user interfaces.

\end{document}